\documentclass[lettersize,journal]{IEEEtran}
\usepackage{amsmath,amsfonts}
\usepackage{algorithmic}
\usepackage{array}
\usepackage[caption=false,font=normalsize,labelfont=sf,textfont=sf]{subfig}
\usepackage{textcomp}
\usepackage{stfloats}
\usepackage{url}
\usepackage{verbatim}
\usepackage{graphicx}
\def\BibTeX{{\rm B\kern-.05em{\sc i\kern-.025em b}\kern-.08em
    T\kern-.1667em\lower.7ex\hbox{E}\kern-.125emX}}
\usepackage{balance}
\usepackage{multirow}
\usepackage{float}
\usepackage{subfig}
\usepackage{enumitem}
\usepackage{booktabs}
\usepackage{algorithm}
\usepackage{algorithmic}
\usepackage{color}
\usepackage{hyperref}
\usepackage{marvosym}
\begin{document}

\newcommand{\shortname}{\emph{GSDA}} 
\newcommand{\fullname}{\textbf{\emph{\underline{G}raph-\underline{S}tructured \underline{D}ual \underline{A}daptation Framework~(GSDA)}}}

\title{Graph-Structured Driven Dual Adaptation for Mitigating Popularity Bias}

\author{Miaomiao Cai, Lei Chen, Yifan Wang, Zhiyong Cheng, Min Zhang~\IEEEmembership{Member,~IEEE}, Meng Wang~\IEEEmembership{Fellow,~IEEE}%
\thanks{This work was supported by the National Key Research and Development Program of China under Grant 2024YFC3307403, and by the National Natural Science Foundation of China under Grants 72188101 and 62272254.}%

\thanks{Miaomiao Cai is with the National University of Singapore, Singapore(\href{mailto:cmm.hfut@gmail.com}{cmm.hfut@gmail.com}). Lei Chen is with the University of Science and Technology of China, Hefei, China(\href{chenlei.hfut@gmail.com}{chenlei.hfut@gmail.com}). Yifan Wang and Min Zhang are with Tsinghua University, Beijing, China(\href{yf-wang21@mails.tsinghua.edu.cn}{yf-wang21@mails.tsinghua.edu.cn},\href{z-m@tsinghua.edu.cn}{z-m@tsinghua.edu.cn}). Zhiyong Cheng and Meng Wang are with the Hefei University of Technology, Hefei, China(\href{mailto:jason.zy.cheng@gmail.com}{jason.zy.cheng@gmail.com}, \href{mailto:eric.mengwang@gmail.com}{eric.mengwang@gmail.com}). (\textit{Corresponding authors: Lei Chen; Meng Wang}.)}
}

\maketitle

\begin{abstract}

Popularity bias is a common challenge in recommender systems. It often causes unbalanced item recommendation performance and intensifies the Matthew effect. Due to limited user-item interactions, unpopular items are frequently constrained to the embedding neighborhoods of only a few users, leading to representation collapse and weakening the model’s generalization. Although existing supervised alignment and reweighting methods can help mitigate this problem, they still face two major limitations: (1) they overlook the inherent variability among different Graph Convolutional Networks(GCNs) layers, which can result in negative gains in deeper layers; (2) they rely heavily on fixed hyperparameters to balance popular and unpopular items, limiting adaptability to diverse data distributions and increasing model complexity.

To address these challenges, we propose ~\fullname, a dual adaptive framework for mitigating popularity bias in recommendation. Our theoretical analysis shows that supervised alignment in GCNs is hindered by the over-smoothing effect, where the distinction between popular and unpopular items diminishes as layers deepen, reducing the effectiveness of alignment at deeper levels. To overcome this limitation, ~\shortname integrates a hierarchical adaptive alignment mechanism that counteracts entropy decay across layers together with a distribution-aware contrastive weighting strategy based on the Gini coefficient, enabling the model to adapt its debiasing strength dynamically without relying on fixed hyperparameters. Extensive experiments on three benchmark datasets demonstrate that ~\shortname effectively alleviates popularity bias while consistently outperforming state-of-the-art methods in recommendation performance. The source code for our method is available at \url{https://github.com/miaomiao-cai2/GSDA}.

\end{abstract}


\section{Introduction}

\IEEEPARstart{R}{ecommender} systems (RS) have become indispensable in various domains, such as e-commerce and streaming services, providing personalized recommendations to alleviate information overload~\cite{covington2016deep, citationsurveylekey, cui2012discover,qiu2019rethinking,dinnissen2022fairness}. Among numerous recommendation paradigms, Collaborative Filtering (CF) remains a cornerstone approach due to its ability to achieve strong recommendation performance by solely leveraging historical user-item interactions. Recently, Graph Convolutional Networks (GCNs) have emerged as prominent methods in CF, effectively capturing high-order collaborative signals through graph structures. However, popularity bias persists as a critical challenge, wherein highly-interacted items dominate recommendation lists, causing unpopular items to be increasingly marginalized~\cite{Chen2020BiasAD, cai2024mitigating,evil}. Such imbalance not only skews recommendation outcomes but also intensifies the "\textbf{Matthew effect}"~\cite{Zhu2021PopularityBI, Wei2020ModelAgnosticCR, abdollahpouri2021user,dinnissen2023amplifying}, further widening the visibility gap between popular and unpopular items.

\begin{figure}[t]
    \centering
    \subfloat{\includegraphics[width =1\linewidth]{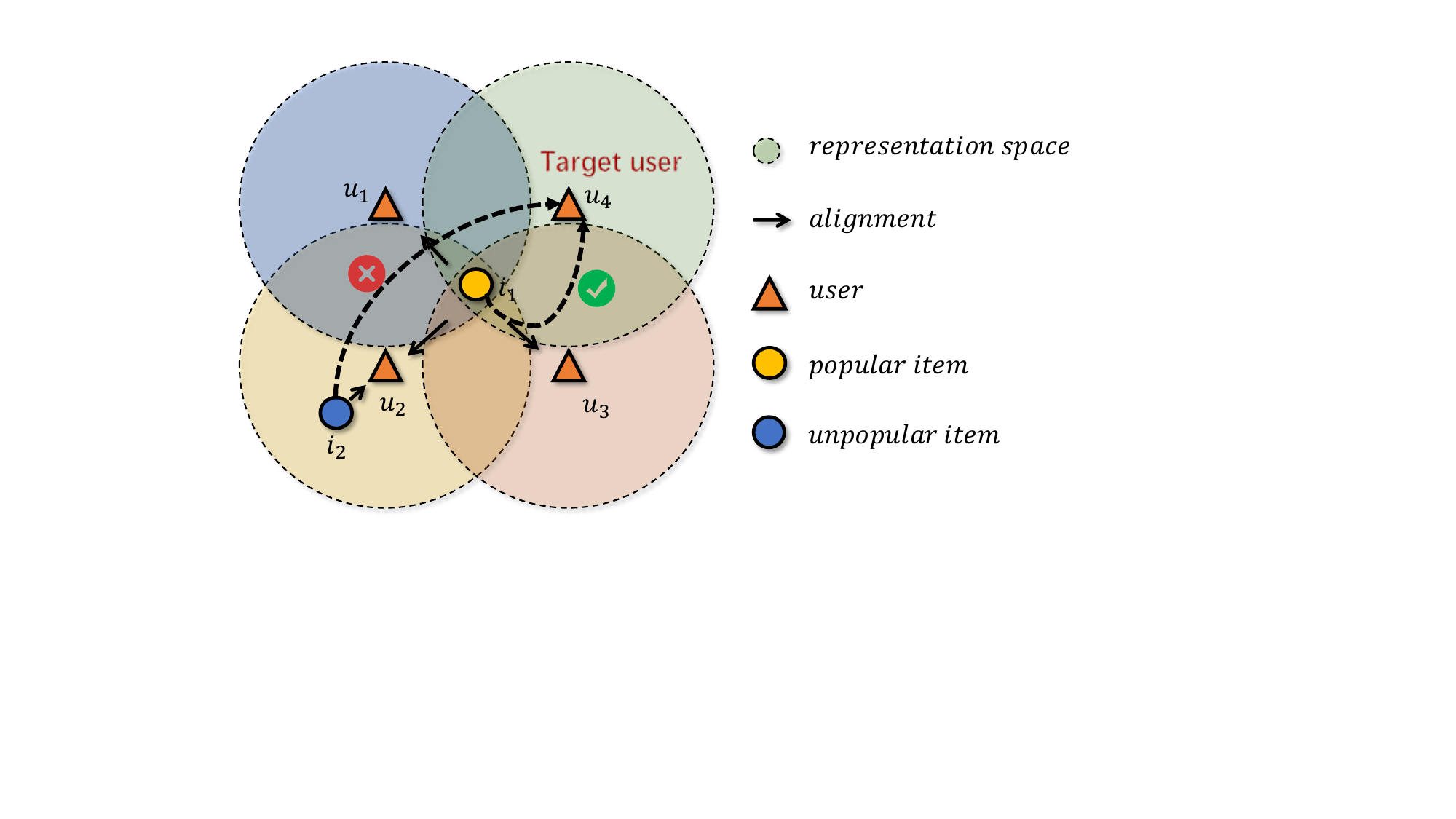}\label{fig:cf_collapse}}
    \caption{Illustration of representation collapse. The popular item ($i_1$) interacts with multiple users ($u_1,u_2,u_3$), obtaining a generalized representation, while the unpopular item ($i_2$), interacting with only one user ($u_2$), suffers from representation collapse due to insufficient contextual information.}
    \label{fig:intro}
\end{figure}

The mechanism of popularity bias is closely tied to the alignment paradigm in Collaborative Filtering (CF), where user and item representations are aligned according to observed user-item interactions~\cite{Wang2022TowardsRA, Rendle2009BPRBP, Koren2009MatrixFT,park2023toward}. However, when these interactions are highly imbalanced, this alignment strategy often results in \textbf{representation collapse} for unpopular items. As illustrated in Fig.\ref{fig:intro}, a target user $u_4$ interacts with both a popular item ($i_1$) and an unpopular item ($i_2$) during the testing phase. During training, the popular item $i_1$ connects to multiple users ($u_1, u_2,$ and $u_3$), enabling it to aggregate richer information and learn a more generalized embedding. In contrast, the unpopular item $i_2$ has minimal exposure, relying excessively on the embedding space of a single user ($u_2$). Consequently, $i_2$ experiences representation collapse due to limited contextual information, further marginalizing unpopular items and exacerbating popularity bias within CF-based recommender systems. This representation collapse fundamentally arises from CF’s supervised alignment paradigm, which relies on user-item interactions for embedding optimization. 

To address representation collapse, existing methods mainly focus on improving embedding distributions (e.g., AURL~\cite{cai2024mitigating}), enhancing embedding uniformity through contrastive learning (e.g., SimGCL~\cite{Yu2021AreGA}, SGL~\cite{Wu2020SelfsupervisedGL}), and employing group-specific weighting or supervised alignment to explicitly differentiate item popularity levels~\cite{cai2024popularityawarealignmentcontrastmitigating,FairnessAware,Useroriented,PopularityawareDRO}. More recently, supervised alignment methods have emerged, explicitly aligning embeddings of popular and unpopular items interacting with the same user~\cite{cai2024popularityawarealignmentcontrastmitigating}. Although these approaches have made considerable progress, they still suffer from critical limitations. 

First, \textbf{existing supervised alignment methods typically overlook the inherent structural differences among layers in Graph Convolutional Networks (GCNs)}. Specifically, shallow layers focus primarily on capturing local neighborhood features, whereas deeper layers aggregate broader, global structural information. Uniformly applying the same alignment constraints across all layers neglects this hierarchy, resulting in ineffective or even counterproductive alignment in deeper layers due to embedding homogenization and semantic drift. Second, \textbf{current group-specific weighting strategies heavily rely on static hyperparameters to balance the contributions of popular and unpopular items}. Such rigid parameter settings require extensive manual tuning for different datasets, limiting model adaptability and scalability in dynamic environments with varying item popularity distributions.

To overcome these limitations, we propose a novel framework called ~\fullname, which explicitly leverages structural and distributional insights derived from the graph's adjacency matrix to effectively mitigate popularity bias. Our theoretical analysis reveals a critical yet previously overlooked issue: as Graph Convolutional Networks (GCNs) deepen, the over-smoothing effect leads to the homogenization of node embeddings, which significantly reduces the conditional entropy between popular and unpopular items. \textbf{This reduction in conditional entropy severely compromises the effectiveness of supervised alignment strategies in deeper layers.} Motivated by this novel theoretical insight, we propose a dual adaptive strategy that uniquely combines layer-wise structural signals and item distribution information from the adjacency matrix. Specifically, our approach ~\shortname ~ includes: (1) a hierarchical adaptive alignment mechanism to address embedding collapse across layers; (2) a distribution-aware dynamic weighting strategy that avoids reliance on fixed hyperparameters by dynamically adapting to changes in item popularity distributions. Extensive experiments validate that our approach effectively mitigates popularity bias and consistently surpasses state-of-the-art baselines in recommendation accuracy.
Our main contributions are summarized as follows:
\begin{itemize}[leftmargin=0.5cm] 
    \item We theoretically and empirically reveal that supervised alignment effectiveness declines significantly in deeper GCN layers due to embedding homogenization (over-smoothing) induced by repeated propagation through higher-order adjacency matrices. 
    \item We introduce a dual adaptive framework that leverages graph structural information to extract item distribution patterns and dynamically adjusts alignment strategies, effectively mitigating popularity bias without extensive hyperparameter tuning. 
    \item Extensive experiments demonstrate our approach’s superior performance, significantly reducing popularity bias while achieving state-of-the-art recommendation accuracy and robustness across diverse scenarios. 
\end{itemize}

\section{Preliminary}
\subsection{Recommendation Task}

Collaborative Filtering (CF), a widely adopted paradigm, predicts these interactions by aligning user and item representations. User-item interactions are represented by a binary matrix $\mathcal{R}\in\{0,1\}^{M\times N}$, where $\mathcal{R}_{u,i}=1$ indicates an interaction between user $u$ and item $i$, and $0$ otherwise. Let $U$ and $I$ denote user and item sets with $|U|=M$ and $|I|=N$, respectively. CF models typically learn user embeddings $\mathbf{Z}\in\mathbb{R}^{M\times D}$ and item embeddings $\mathbf{E}\in\mathbb{R}^{N\times D}$, where $D$ is the embedding dimension. The predicted preference score for user $u$ towards item $i$ is calculated as $s(u,i)=\mathbf{z}_u^\top\mathbf{e}_i$, where $\mathbf{z}_u$ and $\mathbf{e}_i$ are the embeddings for user $u$ and item $i$.

To optimize user and item embeddings, we employ the Bayesian Personalized Ranking (BPR) loss~\cite{Rendle2009BPRBP}, a pairwise ranking objective widely used in recommendation tasks. This loss encourages the model to rank observed (positive) interactions higher than unobserved (negative) interactions:
\begin{equation}
    \mathcal{L}_{rec}= -\frac{1}{|\mathcal{R}|}\sum_{(u,i,j)\in \mathcal{O^+}} ln\sigma(s(u,i)-s(u,j)),
    \label{bprloss}
\end{equation}
where $\sigma(\cdot)$ is the sigmoid function, $\mathcal{O^+}=\{(u, i,j)|\mathcal{R}_{u, i}=1,\mathcal{R}_{u,j}=0\}$ represents pairwise data, and $j$ is a randomly sampled negative item that the user has not interacted with. 

\subsection{Graph Convolutional Networks}
Graph Convolutional Networks (GCNs) have been widely adopted in collaborative filtering~\cite{He2020LightGCNSA,Wang2019NeuralGC,Chen2020RevisitingGB}. Specifically, user-item interactions can be naturally represented as a bipartite graph $\mathcal{G} = (\mathcal{U}\cup\mathcal{I},\mathbf{A})$, where nodes include users and items, and edges represent interactions. The corresponding adjacency matrix $\mathbf{A}\in\mathbb{R}^{(M+N)\times(M+N)}$ is defined as:
\begin{small}
    \begin{flalign}
        \mathbf{A} = \left[\begin{array}{cc}
        \mathbf{0}^{M \times M} & \mathcal{R}\\[6pt]
        \mathcal{R}^\top & \mathbf{0}^{N \times N}
        \end{array}\right],
    \end{flalign}
\end{small}
where $\mathcal{R}\in\{0,1\}^{M\times N}$ denotes the interaction matrix between users and items. To explicitly preserve self-information during the propagation, self-loops are introduced by integrating an identity matrix $\mathbf{I}$, resulting in the adjacency matrix $(\mathbf{A}+\mathbf{I})$. This adjacency matrix is then normalized as follows:
\begin{equation}
    \hat{\mathbf{A}} = \mathbf{D}^{-1/2}\,(\mathbf{A}+\mathbf{I})\,\mathbf{D}^{-1/2},
\end{equation}
where $\mathbf{D}$ is the diagonal degree matrix. And its diagonal elements $\mathbf{D}_{ii}$ represent the degree (sum of connections) of node $i$. This normalized adjacency matrix serves as the basis for propagating embeddings within GCN layers~\cite{yang2023generative}. 

In this work, we employ LightGCN~\cite{He2020LightGCNSA} as the embedding encoder due to its simplicity and strong performance in collaborative filtering tasks. Specifically, we initialize embeddings for all nodes (users and items) as $\mathbf{X}^{(0)} = \bigl[\mathbf{Z}^{(0)}, \mathbf{E}^{(0)}\bigr]$, where $\mathbf{Z}^{(0)}$ and $\mathbf{E}^{(0)}$ denote initial embeddings of users and items, respectively. These embeddings are iteratively updated through a layer-wise propagation rule defined as follows:
\begin{equation}
    \mathbf{X}^{(l)} = \hat{\mathbf{A}} \mathbf{X}^{(l-1)},
\end{equation}
where $\hat{\mathbf{A}}$ denotes the normalized adjacency matrix, and $\mathbf{X}^{(l-1)}$ represents node embeddings from the previous layer. Applying this propagation recursively across $l$ layers, the final embeddings at layer $l$ can be succinctly expressed as:
\begin{equation}
    \mathbf{X}^{(l)} = \hat{\mathbf{A}}^l \mathbf{X}^{(0)}.
    \label{eqa:pro}
\end{equation}
It indicates that the embeddings at layer $l$ are determined by the $l$-th order multiplication of the normalized adjacency matrix, thus capturing multi-hop neighborhood information.

\section{Theoretical and Empirical Validation}
In this subsection, we establish a theoretical connection between supervised alignment effectiveness and conditional entropy. Specifically, we first define conditional entropy as a metric for measuring alignment efficacy (Entropy as a Measure of Alignment Effectiveness). Then, we demonstrate how repeated message passing in GCNs leads to over-smoothing, homogenizing embeddings across layers (Over-Smoothing Caused by High-Order Adjacency Matrices). Finally, we analytically derive the layer-wise reduction in conditional entropy \textbf{due to over-smoothing, showing that it weakens the information transfer capability of supervised alignment at deeper layers} (Layer-Wise Conditional Entropy Reduction). This structured analysis provides a comprehensive understanding of why alignment effectiveness diminishes as GCN depth increases. 

\subsection{Theoretical Analyses}
\textbf{Entropy as a Measure of Alignment Effectiveness.} Supervised alignment aims to mitigate popularity bias by explicitly aligning embeddings of popular and unpopular items that share interactions with the same user~\cite{cai2024popularityawarealignmentcontrastmitigating}. To quantitatively evaluate the effectiveness of supervised alignment, we leverage an information-theoretic~\cite{wang2003new, zhang2023mitigating} perspective by defining a metric based on entropy. Specifically, we measure how well embeddings of unpopular items retain independent information when conditioned on popular item embeddings. This motivates the use of conditional entropy as a direct indicator of supervised alignment performance.
\begin{equation}
    H(\mathbf{X}_{\text{up}}|\mathbf{X}_{\text{p}}) = -\mathbb{E}_{P(\mathbf{X}_{\text{up}},\mathbf{X}_{\text{p}})}[\log P(\mathbf{X}_{\text{up}}|\mathbf{X}_{\text{p}})],
    \label{eq:cond_entropy_def}
\end{equation}
where $\mathbf{X}_{\text{up}}$ and $\mathbf{X}_{\text{p}}$ denote embeddings of unpopular and popular items, respectively. This entropy-based metric directly reflects alignment efficacy—higher values indicate more effective information transfer. In particular, a higher conditional entropy indicates greater uncertainty about unpopular item embeddings, thus implying more potential for beneficial information transfer from popular to unpopular items. Conversely, lower conditional entropy implies that embeddings of unpopular items already possess similar information to popular items, limiting alignment benefits.
For analytical tractability, we further assume that initial node embeddings are sampled independently from an isotropic Gaussian distribution with zero mean, which allows expected squared distances across nodes to be reduced to a constant factor. This standard assumption in GCN theory simplifies the derivation without affecting the qualitative conclusions about entropy decay.

\textbf{Over-Smoothing Caused by High-Order Adjacency Matrices.} 
Having established entropy as a key metric for alignment effectiveness, we now analyze how GCN message passing affects entropy. In particular, we show that repeated propagation through the normalized adjacency matrix leads to over-smoothing, which directly impacts entropy-based alignment performance. In Graph Convolutional Networks (GCNs), node embeddings at the \(l\)-th layer are generated by recursively propagating information through the normalized adjacency matrix \(\hat{\mathbf{A}}\). Formally, this process is defined as Eq.~\eqref{eqa:pro}, where \(\mathbf{X}^{(0)}\) represents initial node embeddings, typically initialized from a random Gaussian distribution with zero mean, and \(\hat{\mathbf{A}}\) is the normalized adjacency matrix encoding structural relationships among nodes.

As the depth \(l\) increases, the repeated multiplication of \(\hat{\mathbf{A}}\) leads its entries to become progressively uniform~\cite{Zhao2022InvestigatingAP}. Specifically, the differences between any two entries in the high-order adjacency matrix diminish, satisfying: 
\begin{equation} 
    \lim_{l \to \infty} \left|\hat{\mathbf{A}}^l_{u,j} -\hat{\mathbf{A}}^l_{p,k}\right| = 0, \quad \text{for all nodes } u, p, j, k. 
    \label{eq:mat_uniform}
\end{equation}

This convergence requires the user–item graph to be connected and aperiodic; in practice, LightGCN satisfies this condition by including self-loops. This convergence means that, as layers deepen, the propagation weights between any two nodes become almost indistinguishable. When combined with the random initialization of embeddings, this uniformity drives different nodes to become increasingly similar, a phenomenon widely known as over-smoothing. For unpopular items with sparse connections, this loss of distinctiveness is particularly severe.

\textbf{Layer-Wise Conditional Entropy Reduction.}
Given the homogenization of embeddings induced by over-smoothing, we now analyze how this phenomenon directly leads to a monotonic reduction in conditional entropy. By explicitly deriving the relationship between adjacency matrix differences and entropy, we demonstrate that alignment effectiveness inevitably weakens as the number of layers increases. We now explicitly connect the over-smoothing phenomenon described above to the progressive reduction in conditional entropy observed in supervised alignment. Consider the embeddings of an unpopular item \(\mathbf{X}_{\text{up}}^{(l)}\) and a popular item \(\mathbf{X}_{\text{p}}^{(l)}\) at layer \(l\), which can be expressed based on GCN propagation: 
 \begin{equation} 
    \mathbf{X}_{\text{up}}^{(l)} = \sum_{j}\hat{\mathbf{A}}^l_{up,j}\mathbf{x}_j^{(0)},\quad \mathbf{X}_{\text{p}}^{(l)} = \sum_{j}\hat{\mathbf{A}}^l_{p,j}\mathbf{x}_j^{(0)},
    \label{eq:xp_up_layer} 
\end{equation} 
where \(\hat{\mathbf{A}}^l_{up,j}\) and \(\hat{\mathbf{A}}^l_{p,j}\) represent entries in the normalized adjacency matrix at layer \(l\), quantifying how information propagates from node \(j\) to the unpopular item \(up\) and popular item \(p\), respectively. Here, \(\mathbf{x}_j^{(0)}\) denotes the initial embedding vector of node \(j\), initialized from a random Gaussian distribution.

To quantify how well popular item embeddings can inform unpopular item embeddings, we use conditional entropy as defined Eq.~\eqref{eq:cond_entropy_def}. For tractable analysis, we approximate the conditional probability \(P(\mathbf{X}_{\text{up}}^{(l)} \mid \mathbf{X}_{\text{p}}^{(l)})\) via Gaussian kernel density estimation (KDE)~\cite{botev2010kernel}: 
\begin{equation} 
    P(\mathbf{X}_{\text{up}}^{(l)} \mid \mathbf{X}_{\text{p}}^{(l)})\;\propto\;\exp\Bigl(-\frac{\|\mathbf{X}_{\text{up}}^{(l)} - \mathbf{X}_{\text{p}}^{(l)}\|^2}{2\sigma^2}\Bigr), 
    \label{eq:kde_cond} 
\end{equation}
where \(\sigma\) is a kernel bandwidth parameter controlling the smoothness of the density estimation. This KDE surrogate is used solely to relate conditional entropy to squared Euclidean distance. The constant terms introduced by this approximation are independent of the layer index $l$ and therefore do not affect the monotonic trend we establish.
By substituting Eq.~\eqref{eq:kde_cond} into the conditional entropy definition, we explicitly establish a relationship between entropy and embedding distance: 
\begin{equation}    
    H(\mathbf{X}_{\text{up}}^{(l)}\mid\mathbf{X}_{\text{p}}^{(l)})\;\propto\;\mathbb{E}\bigl[\|\mathbf{X}_{\text{up}}^{(l)} - \mathbf{X}_{\text{p}}^{(l)}\|^2\bigr]. 
    \label{eq:entropy_distance} 
\end{equation}
Expanding the squared Euclidean distance term explicitly using the definitions in Eq.~\eqref{eq:xp_up_layer} yields: 
\begin{equation} 
    \|\mathbf{X}_{\text{up}}^{(l)} - \mathbf{X}_{\text{p}}^{(l)}\|^2 \;=\; \sum_{j,k}\,\bigl(\hat{\mathbf{A}}^l_{up,j} - \hat{\mathbf{A}}^l_{p,k}\bigr)^2\,\|\mathbf{x}_j^{(0)} - \mathbf{x}_k^{(0)}\|^2. 
    \label{eq:expand_dist} 
\end{equation} 
This step explicitly clarifies how the differences between adjacency matrix entries directly determine embedding distances. Taking expectation over the initial embeddings (assuming Gaussian initialization with zero mean), the conditional entropy can be approximated as: 
\begin{equation} 
\small
    H\bigl(\mathbf{X}_{\text{up}}^{(l)} \mid \mathbf{X}_{\text{p}}^{(l)}\bigr)\;\propto\;\sum_{j,k}\bigl(\hat{\mathbf{A}}^l_{up,j} - \hat{\mathbf{A}}^l_{p,k}\bigr)^2\mathbb{E}\bigl[\|\mathbf{x}_j^{(0)} - \mathbf{x}_k^{(0)}\|^2\bigr].     \label{eq:final_entropy}
\end{equation}

This entropy decreases monotonically with layer depth $l$, reflecting the diminishing distinctiveness of unpopular items. As the number of layers increases, over-smoothing makes adjacency entries between popular and unpopular items progressively similar. This shrinking difference reduces embedding distances, which in turn leads to a monotonic decline in conditional entropy across layers. Put simply, deeper propagation erodes the unique information of unpopular items, weakening the benefit of alignment. Unlike prior over-smoothing analyses~\cite{li2018deeper, Oono2021Expressivity, liu2020towards}, which treat all nodes symmetrically, our analysis explicitly incorporates the degree heterogeneity of user–item bipartite graphs. 
We show that high-degree (popular) items dominate the leading spectral components of $\hat{\mathbf{A}}$, causing low-degree (long-tail) items to collapse towards them. 
This provides a new perspective: over-smoothing is not only a homogenization phenomenon but also a mechanism that amplifies popularity bias.
\begin{equation} 
    H\bigl(\mathbf{X}_{\text{up}}^{(l)}\mid\mathbf{x}_p^{(l)}\bigr) \;\ge\; H\bigl(\mathbf{X}_{\text{up}}^{(l+1)}\mid\mathbf{x}_p^{(l+1)}\bigr). 
    \label{eq:monotonic_decay} 
\end{equation}

In summary, repeated propagation through higher-order adjacency matrices homogenizes embeddings, progressively reducing conditional entropy and weakening the effectiveness of supervised alignment in deeper GCN layers.

\subsection{Experimental Validation}

\begin{figure}[t]
    \centering
    \subfloat[\scriptsize Conditional Entropy and Similarity]{\includegraphics[width =0.5\linewidth]{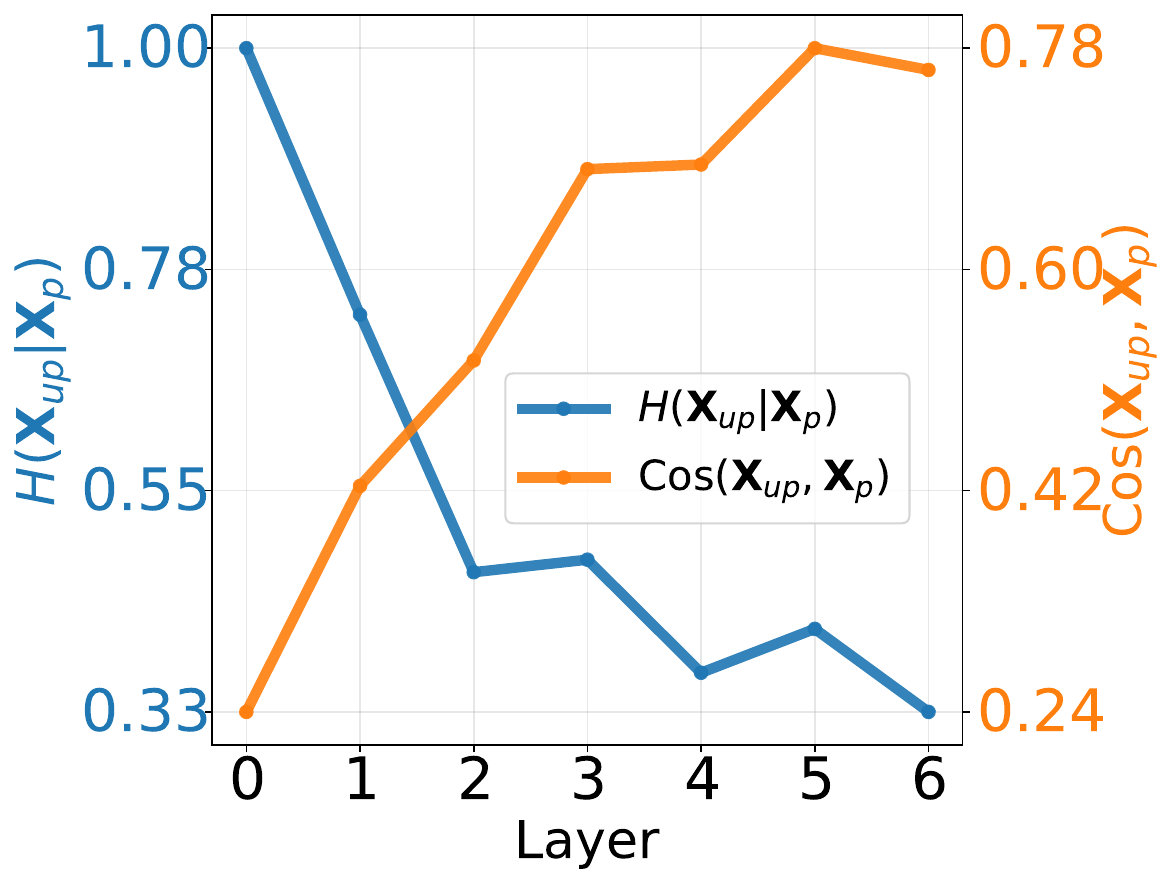}\label{fig:intro_ces}}
    \hfill
    \subfloat[\scriptsize Supervised Alignment]{\includegraphics[width =0.5\linewidth]{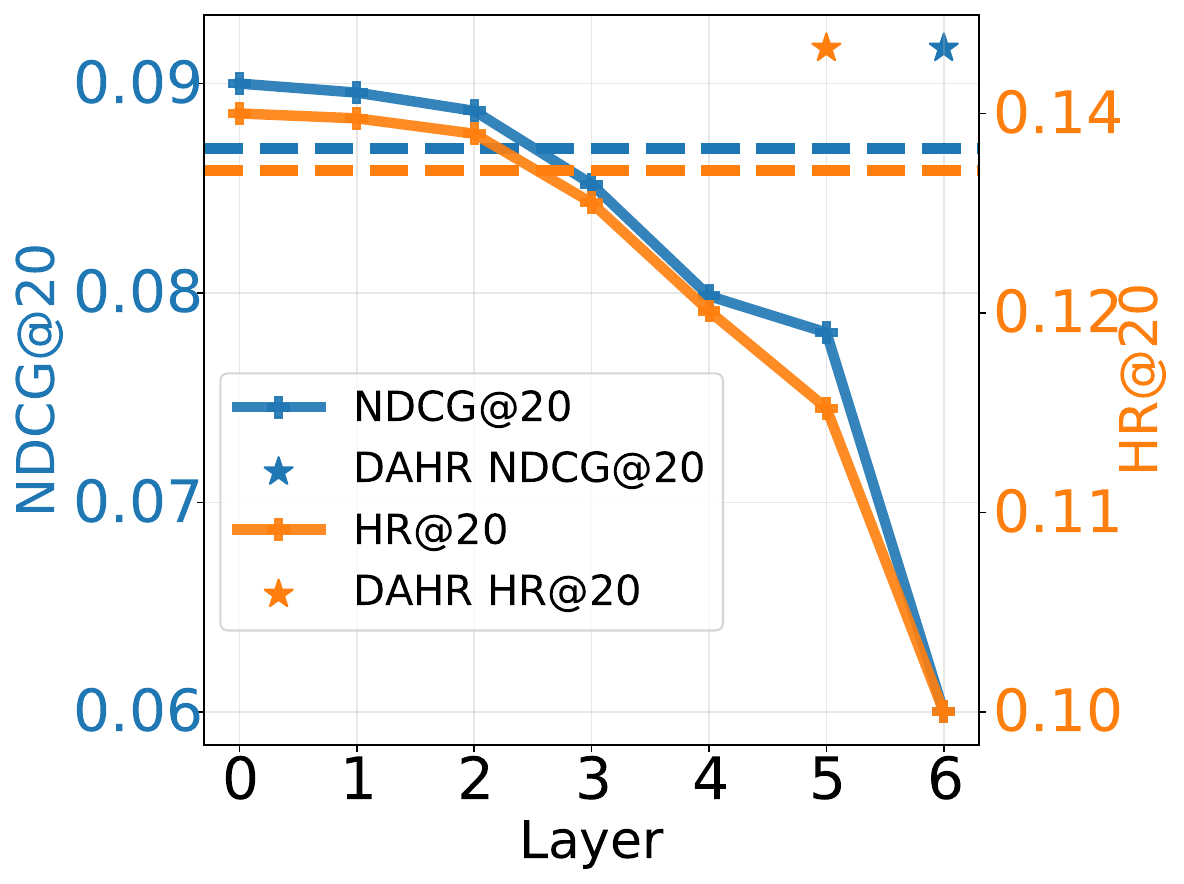}\label{fig:intro_metrics}}
    \caption{((Left) Variation in conditional entropy and embedding similarity across GCN layers, illustrating how deeper layers lead to increased homogenization. (Right) Effect of applying supervised alignment at different GCN depths, highlighting its impact on recommendation performance and the diminishing gains at deeper layers.}
    \label{fig:intro_sec_2}
\end{figure}

We perform experiments on the Gowalla dataset to empirically examine the relationship among over-smoothing, embedding homogenization, and supervised alignment effectiveness in deeper layers. First, we randomly sample 1,000 pairs of popular and unpopular items and calculate their conditional entropy ($H(\mathbf{X}_{\text{up}}^{(l)}|\mathbf{X}_p^{(l)})$) and cosine similarity across different layers of LightGCN. As shown in Fig.\ref{fig:intro_ces}, the conditional entropy between popular and unpopular items significantly decreases as the layer depth increases, coinciding with increased embedding similarity. This observation empirically confirms our theoretical finding that *embedding homogenization occurs due to the progressive uniformity of the adjacency matrix entries in deeper layers (Eq.~\eqref{eq:mat_uniform}), which directly reduces the distinguishability of unpopular item embeddings.

Furthermore, we evaluate the effectiveness of supervised alignment using the PAAC method~\cite{cai2024popularityawarealignmentcontrastmitigating} applied at various layer depths on Gowalla, as illustrated in Fig.\ref{fig:intro_metrics}. Performance metrics (e.g., NDCG@20, Recall@20) clearly demonstrate that supervised alignment applied at shallow layers significantly improves recommendation quality, while alignment in deeper layers leads to negative gains. This aligns with our theoretical derivation that conditional entropy decreases monotonically with depth (Eq.~\eqref{eq:monotonic_decay}), leading to diminishing information transfer potential from popular to unpopular items. Specifically, as shown in Eq.~\eqref{eq:final_entropy}, the reduced variance in adjacency matrix weights across nodes causes the expected embedding distance to shrink, directly driving the observed entropy decay.

In summary, both theoretical analysis and empirical validation underline the critical limitation of existing supervised alignment approaches: \textbf{repeated propagation via higher-order adjacency matrices leads to embedding homogenization}(Eq.~\eqref{eq:mat_uniform}), \textbf{which progressively reduces conditional entropy} (Eq.~\eqref{eq:monotonic_decay}) and \textbf{thus weakens the effectiveness of supervised alignment at deeper GCN layers}. These findings strongly support the necessity of developing adaptive strategies that adjust alignment strength based on layer depth, as explored in our proposed dual adaptive strategy.

\section{Methodology}

\begin{figure*}[t]
    \centering
    \includegraphics[width=1\linewidth]{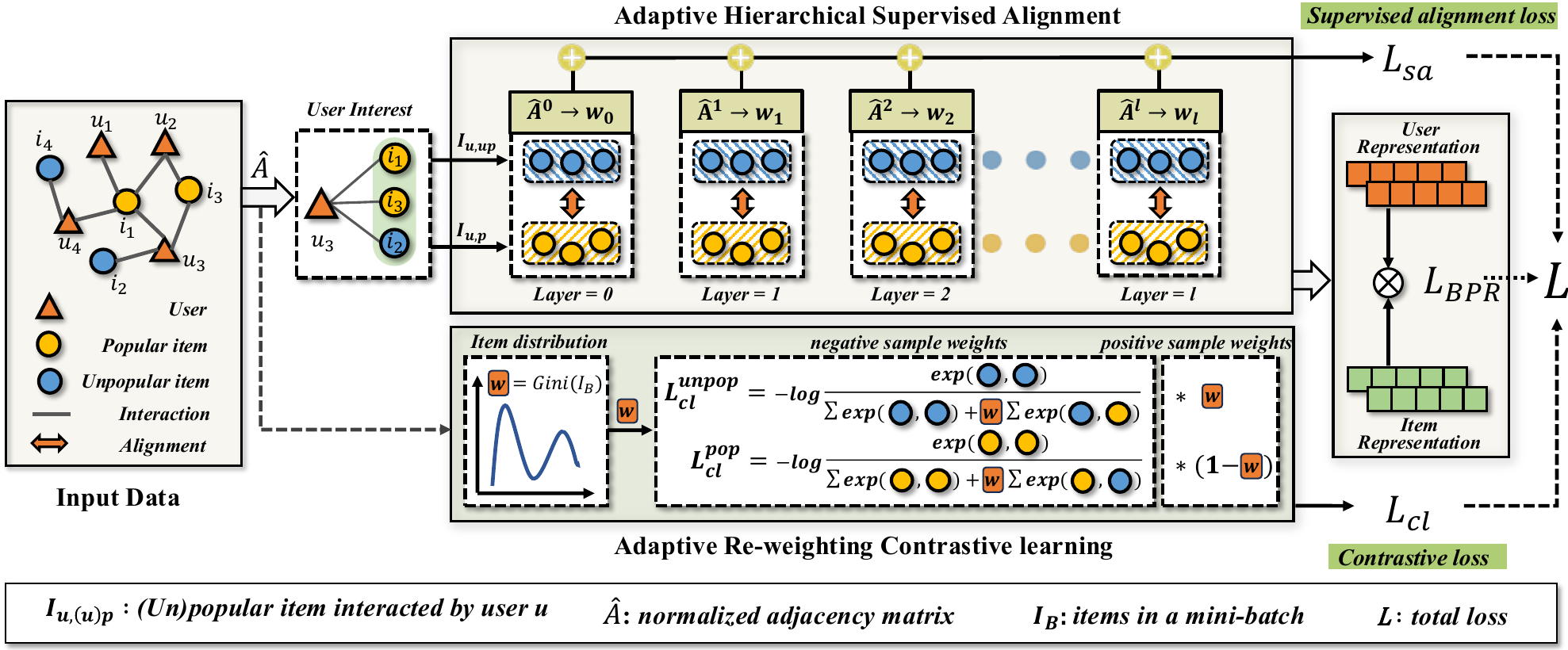}
    \caption{The \fullname\  consists of two key modules: (1) \textbf{Adaptive Hierarchical Supervised Alignment}, which reduces over-smoothing in deeper GCN layers; and (2) \textbf{Adaptive Re-weighting Contrastive Learning}, which dynamically adjusts sample weights based on real-time popularity distributions. }
    \label{fig:framework}
\end{figure*} 

To address the limitations identified earlier, we propose a novel dual-adaptive framework, termed ~\fullname, which explicitly leverages the structural and distributional properties encoded within the graph's high-order adjacency matrix to mitigate popularity bias. Our theoretical and empirical analyses uncover a critical yet previously overlooked limitation: as Graph Convolutional Networks (GCNs) deepen, embeddings derived from high-order adjacency matrices ($\hat{\mathbf{A}}^l$) progressively become more uniform due to the over-smoothing effect. This increasing uniformity results in significant homogenization of node embeddings, thereby reducing the conditional entropy between popular and unpopular item representations. Consequently, the effectiveness of supervised alignment strategies diminishes, particularly at deeper layers.

Inspired by these insights, we propose ~\fullname, which leverages structural features and distribution information embedded in the high-order adjacency matrix to address these challenges. The specific framework model diagram is shown in Fig.\ref{fig:framework}. Specifically, we introduce two innovative modules: (1) Adaptive Hierarchical Supervised Alignment, which employs a carefully designed adaptive weight decay mechanism based on structural information from higher-order adjacency matrices to alleviate the weakening effects of supervised alignment in deeper layers; and (2) Adaptive Re-weighting Contrastive Learning, which dynamically estimates item popularity distributions utilizing adjacency matrix information, adaptively adjusting embedding weights in real-time. This eliminates the dependency on fixed hyperparameters, enhancing the model's adaptability and robustness across varying item distributions.

\subsection{Adaptive Hierarchical Supervised Alignment}

Our theoretical and empirical analyses have clearly demonstrated that the effectiveness of supervised alignment between popular and unpopular items deteriorates significantly with increasing Graph Convolutional Network (GCN) layer depth. The fundamental reason behind this issue is the homogenization of node embeddings in deeper layers, caused by the over-smoothing effect when repeatedly propagating embeddings via the high-order normalized adjacency matrix ($\hat{\mathbf{A}}^l$). Specifically, as the layer depth $l$ increases, embeddings become increasingly uniform, reducing the conditional entropy between popular and unpopular item representations. This entropy reduction directly limits the ability of supervised alignment to transfer useful information from popular items to unpopular ones, ultimately weakening its effectiveness in mitigating popularity bias.

To address this issue, we propose the Adaptive Hierarchical Supervised Alignment module, which dynamically adjusts the alignment influence across different GCN layers based on structural characteristics encoded in the adjacency matrix. Following the embedding propagation mechanism used in LightGCN~\cite{He2020LightGCNSA}, node embeddings at layer $l$ are computed as:
\begin{equation}
\mathbf{X}^{(l)} = \hat{\mathbf{A}}^l \mathbf{X}^{(0)},
\label{eqa:propgation}
\end{equation}
where $\mathbf{X}^{(0)}$ represents the initial node embeddings, and $\hat{\mathbf{A}}$ denotes the normalized adjacency matrix. At the base layer ($l=0$), the adjacency matrix is defined as the identity matrix, i.e., $\mathbf{A}^0 = \mathbf{I}$, ensuring the embeddings at layer 0 solely reflect initial self-information:
\begin{equation}
\mathbf{X}^{(0)} = \mathbf{I} \mathbf{X}^{(0)} = \mathbf{X}^{(0)}.
\end{equation}

Given our theoretical insight that deeper layers exhibit increased embedding uniformity, we introduce an adaptive layer-wise weighting strategy informed by the Frobenius norm of adjacency matrix powers:
\begin{equation}
w_l = \frac{\|\hat{\mathbf{A}}^l\|_F}{\sum_{k=0}^{L}\|\hat{\mathbf{A}}^k\|_F},
\label{eqa:wl}
\end{equation}
where $\|\hat{\mathbf{A}}^l\|_F$ represents the Frobenius norm of the adjacency matrix at layer $l$, quantifying the diversity or richness of structural information retained after $l$-hop propagation. The Frobenius norm effectively measures how distinctively the adjacency structure is preserved at each propagation step. As $l$ increases, the entries in $\hat{\mathbf{A}}^l$ tend to become more uniform due to over-smoothing, leading to a smaller Frobenius norm. Consequently, this adaptive weighting strategy inherently assigns greater emphasis to shallow layers (which preserve richer localized information) and progressively reduces weights at deeper layers.

Given the adaptive layer-wise weighting scheme, the hierarchical supervised alignment loss across layers is defined as:
\begin{equation}
    \small
    \mathcal{L}_{sa}=\frac{1}{L+1}\sum_{l=0}^{L} w_l \left(\frac{1}{|U|}\sum_{u \in \mathcal{U}} \frac{1}{|\boldsymbol{I}_u|}\sum_{i \in \boldsymbol{I}_{u, \text{up}}, j \in \boldsymbol{I}_{u, \text{p}}} \|\mathbf{x}_{i}^{(l)} - \mathbf{x}_{j}^{(l)}\|^2\right),
    \label{eqa:sa}
\end{equation}
where $U$ denotes the set of users, and $\boldsymbol{I}_{u, \text{p}}$ and $\boldsymbol{I}_{u, \text{up}}$ are the sets of popular and unpopular items interacted with by user $u$, respectively. The term $\|\mathbf{x}_{i}^{(l)} - \mathbf{x}_{j}^{(l)}\|^2$ explicitly aligns embeddings of popular and unpopular items at each layer, with weights adaptively adjusted to reflect the structural significance captured at different depths.

By emphasizing shallow-layer embeddings, where conditional entropy and representational distinctiveness remain relatively high, our adaptive strategy preserves the beneficial effects of supervised alignment. At deeper layers, where embeddings suffer more pronounced homogenization, it systematically reduces alignment weights. This design allows ~\shortname  ~ to dynamically adjust alignment contributions based on structural insights derived directly from the graph. As a result, it enhances robustness against popularity bias and improves recommendation performance.

\subsection{Adaptive Re-weighting Contrastive Learning}
Existing contrastive learning methods in recommendation typically use fixed hyperparameters to balance the influence between popular and unpopular items, which restricts their adaptability and effectiveness across different datasets and dynamic environments. Furthermore, these static strategies neglect the nuanced variations in item popularity distributions, leading to potential misrepresentations and suboptimal performance. To explicitly overcome these limitations, we propose an adaptive re-weighting contrastive learning module that dynamically adjusts the weights of samples based on real-time item popularity distributions extracted directly from structural characteristics of the graph's adjacency matrix.

Specifically, rather than assigning uniform or fixed weights, we introduce a user-specific weighting strategy that more accurately captures each user's contribution to item popularity. Each user \( u \) is assigned a weight inversely proportional to the logarithm of their degree (number of interactions), formally defined as:
\begin{equation}
    \label{equa:w_u}
    w_u = \frac{1}{\log(1+d_u) + \varepsilon},
\end{equation}
where \( d_u \) is the degree of user \( u \), and \( \varepsilon \) is a small constant introduced to maintain numerical stability. This log-based formulation provides a smoother penalty for highly active users compared to the linear inverse design, thereby avoiding excessive down-weighting. By adopting this user-specific weighting, the influence of frequent users is moderated in a more balanced manner, leading to a more stable and nuanced estimation of item popularity.

Based on this adjusted weighting, we then derive the item popularity \( p_i^{\ast} \) by aggregating interactions between users and items, weighted by the user-specific factors:
\begin{equation}
    \label{item_pop}
    p_i^{\ast} = \sum_{u \in \mathcal{U}} w_u \cdot \widetilde{A}_{u, i+M},
\end{equation}
where \( \widetilde{A}_{u, i+U} \) represents the normalized interaction strength between user \( u \) and item \( i \). This formulation provides a refined and dynamic estimation of item popularity, effectively reflecting the true underlying distributional dynamics captured from the graph structure.

With this dynamically computed item popularity distribution, we quantify the level of imbalance between popular and unpopular items using the Gini coefficient, defined as:
\begin{equation}
    \label{eq:gini}
    w = \frac{1}{2 I \overline{p^{\ast}}} \sum_{i=0}^{I-1} \sum_{j=0}^{I-1} \left| p_i^{\ast} - p_j^{\ast} \right|,
\end{equation}
where \( \overline{p^{\ast}} \) denotes the average weighted popularity of all items. The Gini coefficient measures the inequality or imbalance in the distribution of item popularity, with a higher value indicating a larger imbalance (i.e., few items dominate interactions).

Finally, we integrate this dynamic measurement into contrastive learning framework, adaptively adjusting the weights for popular and unpopular items based on the imbalance captured by the Gini coefficient (Eq.~\eqref{eq:gini}). Specifically, we propose a variant of contrastive loss that dynamically modulates the influence of popular and unpopular items when used as positive or negative samples:
\begin{equation}
    \mathcal{L}_{cl} = (1-w) \mathcal{L}_{cl}^{\text{p}} + w \mathcal{L}_{cl}^{\text{up}},
    \label{eqa:cl}
\end{equation}
where \( \mathcal{L}_{cl}^{\text{p}} \) and \( \mathcal{L}_{cl}^{\text{up}} \) represent the contrastive losses computed separately for popular and unpopular items. By adaptively adjusting weights based on \( w \), our method dynamically reduces the dominance of the over-represented group. When the distribution is highly skewed (large \( w \)), the model strategically reduces the influence of popular items to enhance embedding diversity and robustness. Conversely, when item popularity is relatively balanced (small \( w \)), the method treats both groups more evenly, maintaining optimal performance.

This adaptive design is driven entirely by real-time insights from the high-order adjacency matrix, eliminating reliance on fixed hyperparameters. By dynamically capturing nuanced graph-based structural signals, our approach ensures effective adaptation to varying item distributions, thereby significantly improving the robustness and generalizability of recommendations while effectively mitigating popularity bias.
\subsection{Model Optimization and Algorithm}

To effectively mitigate the two main limitations of popularity bias, we employ a multi-task training approach~\cite{Wu2020SelfsupervisedGL,wang2023unbiased}. The overall loss function for ~\shortname ~ is defined as follows:
\begin{equation} 
    \label{overall_equa} 
    \mathcal{L}=\mathcal{L}_{rec} \;+\;\lambda_{1}\,\mathcal{L}_{sa}\;+\;\lambda_{2}\,\mathcal{L}_{cl}, 
\end{equation} 
where $\mathcal{L}_{rec}$ is the base recommendation loss, capturing fundamental user–item preference signals (e.g., BPR~\cite{Rendle2009BPRBP} or another ranking objective); $\mathcal{L}_{sa}$ is the adaptive hierarchical supervised alignment loss, designed to address the over-smoothing effect in deeper GCN layers and preserve discriminative item representations;$\mathcal{L}_{cl}$ is the adaptive re-weighting contrastive loss, derived from dynamic item popularity estimation, enabling the model to flexibly adjust sample weights based on real-time popularity distributions. The hyperparameters $\lambda_{1}$ and $\lambda_{2}$ control the relative importance of $\mathcal{L}_{sa}$ and $\mathcal{L}_{cl}$, respectively. After training, we perform inference by computing the dot product between the final user and item embeddings, ranking items based on their predicted relevance~\cite{yang2023generative}. The detailed algorithm for ~\shortname~ is presented in Algorithm~\ref{alg:shortname}. For reproducibility, the source code of our implementation is publicly available at \url{https://github.com/miaomiao-cai2/GSDA}.

Our theoretical and experimental findings have demonstrated that over-smoothing in deeper GCN layers—caused by high-order adjacency matrices becoming increasingly uniform—significantly undermines the effectiveness of supervised alignment. Meanwhile, fixed hyperparameter schemes in contrastive learning fail to accommodate diverse item popularity distributions, resulting in suboptimal performance. By integrating these insights into a graph-structured, dual-adaptive framework, \shortname\ systematically addresses both issues. First, by \emph{adapting layer-wise alignment weights to the structure of higher-order adjacency matrices}, we reduce the detrimental impact of over-smoothing on supervised alignment. Second, by \emph{dynamically estimating item popularity from real-time adjacency matrix signals}, we remove the reliance on static hyperparameters in contrastive learning.  Through this synergy, \shortname\ not only boosts recommendation accuracy but also adapts robustly to varying popularity distributions across datasets. In doing so, it significantly alleviates the “\textbf{rich-get-richer}” phenomenon pervasive in recommender systems, offering a powerful and flexible solution to popularity bias.

\begin{algorithm}[t]
\renewcommand{\algorithmicrequire}{\textbf{Input:}}
\renewcommand{\algorithmicensure}{\textbf{Output:}}
\caption{Training Procedure of \shortname}
\label{alg:shortname}
\begin{algorithmic}[1]
\REQUIRE 
  Interaction matrix $\mathcal{R}$, 
  Normalized adjacency matrix $\widetilde{\mathbf{A}}$,
  Learning rate $\eta$, 
  Hyperparameters $\lambda_1, \lambda_2$,
  Layer depth $L$,
  Stability constant $\varepsilon$
\ENSURE Trained model parameters $\Theta$

\STATE Initialize $\Theta$ (user/item embeddings and GCN weights)
\FOR{epoch $= 1$ \TO $T$}
  \FOR{mini-batch $\mathcal{B} \subseteq \mathcal{R}$}
    \STATE \textbf{Embedding Propagation:}
    \FOR{$l = 0$ \TO $L$} 
      \STATE $\mathbf{X}^{(l)} \leftarrow \widetilde{\mathbf{A}}^l \mathbf{X}^{(0)}$  \hfill \textit{(Eq.\,\ref{eqa:propgation})}
    \ENDFOR
    
    \STATE \textbf{Adaptive Hierarchical Alignment:}
    \FOR{$l = 0$ \TO $L$}
      \STATE $w_l \leftarrow \|\widetilde{\mathbf{A}}^l\|_F / \sum_{k=0}^L \|\widetilde{\mathbf{A}}^k\|_F$ \hfill \textit{(Eq.\,\ref{eqa:wl})}
    \ENDFOR
    \STATE Compute $\mathcal{L}_{sa}$ \hfill \textit{(Eq.~\ref{eqa:sa} )} 

    \STATE \textbf{Adaptive Contrastive Reweighting:}
    \FOR{each user $u \in \mathcal{B}$}
      \STATE $w_u \leftarrow 1/(d_u + \varepsilon)$  \hfill \textit{(Eq.~\ref{equa:w_u} )} 
    \ENDFOR
    \STATE Compute $p_i^* \leftarrow \sum_u w_u \widetilde{A}_{u,i+U}$  \hfill \textit{(Eq.~\ref{item_pop} )} 
    \STATE Compute $w \leftarrow \text{Gini}(\{p_0^*,...,p_{I-1}^*\})$ \hfill \textit{(Eq.~\ref{eq:gini} )} 
    \STATE Compute $\mathcal{L}_{cl} \leftarrow (1-w)\mathcal{L}_{cl}^p + w\mathcal{L}_{cl}^{\text{up}}$  \hfill \textit{(Eq.~\ref{eqa:cl} )} 

    \STATE \textbf{Multi-task Optimization:}
    \STATE $\mathcal{L} \leftarrow \mathcal{L}_{rec} + \lambda_1\mathcal{L}_{sa} + \lambda_2\mathcal{L}_{cl}$  \hfill \textit{(Eq.~\ref{overall_equa} )} 
    \STATE $\Theta \leftarrow \Theta - \eta\nabla_\Theta\mathcal{L}$ \COMMENT{Parameter update}
    
    \IF{validation performance plateaus}
      \STATE \textbf{break} \COMMENT{Early stopping}
    \ENDIF
  \ENDFOR
\ENDFOR
\end{algorithmic}
\end{algorithm}

\subsection{Model Analysis}
\subsubsection{Space Complexity}
GSDA follows LightGCN in maintaining user and item embeddings, requiring $(M+N)d$ parameters, where $M$ and $N$ are the numbers of users and items, and $d$ is the embedding dimension. During training, the adaptive hierarchical supervised alignment module stores intermediate embeddings for $L$ layers, incurring $\mathcal{O}(L(M+N)d)$ additional memory. Since $L$ is typically small, this overhead is modest. The adaptive re-weighting contrastive learning module only stores lightweight popularity statistics (e.g., user degrees, item popularity scores, Gini coefficient), which are negligible compared to embedding storage.

\subsubsection{Time Complexity}
The main computational cost of GSDA arises from the sparse embedding propagation in $L$ layers, identical to LightGCN, with complexity $\mathcal{O}(|R^{+}|Ld)$, where $|R^{+}|$ is the number of non-zero entries in the interaction matrix. The adaptive hierarchical supervised alignment adds $\mathcal{O}(L|\mathcal{B}|d)$ cost for aligning popular–unpopular item pairs in a minibatch $\mathcal{B}$; in practice, a small constant number of pairs per user is sampled, so this remains linear in the batch size. The adaptive re-weighting contrastive learning introduces a similar $\mathcal{O}(L|\mathcal{B}|d)$ term for similarity computation. Popularity statistics and the Gini coefficient are computed once per epoch with negligible cost. Therefore, the overall per-epoch complexity is $\mathcal{O}(|R^{+}|Ld + L|\mathcal{B}|d)$, which is comparable to existing GCN-based debiasing methods and practical in real-world settings.

\subsubsection{Limitations}
Although GSDA achieves consistent improvements in debiasing and recommendation accuracy, its performance can be influenced by certain data characteristics. The adaptive hierarchical alignment assumes that the user–item graph provides sufficiently informative structural signals; in extremely sparse or highly irregular graphs, these signals may be weak, reducing alignment effectiveness. The adaptive re-weighting contrastive learning relies on accurate popularity estimation, and in highly dynamic environments with abrupt popularity shifts, frequent re-estimation may introduce transient instability and extra computation. Moreover, GSDA is designed for bipartite collaborative filtering graphs without side information; in cold-start or heterogeneous scenarios, incorporating additional modalities or graph types may be necessary to maintain robustness. Future work could explore temporal smoothing of popularity estimation, efficient incremental graph updates, and extensions to multi-relational or multimodal settings.

\section{Experiments}

\subsection{Experimental Settings}

\subsubsection{Datasets} 
We conducted experiments on three widely-used recommendation benchmarks: Gowalla\footnote{http://snap.stanford.edu/data/loc-gowalla.html}, Movielens-10M\footnote{https://grouplens.org/datasets/movielens/}, and Globo\footnote{https://paperswithcode.com/dataset/news-interactions-on-globo-com}. Gowalla is a location-based social networking dataset that captures sparse user-item interactions with a pronounced long-tail item distribution, ideal for evaluating popularity bias mitigation. Movielens-10M is a popular dataset comprising 10 million user-movie ratings, which we converted from explicit feedback to implicit feedback for consistency in our experiments. Globo is a news recommendation dataset capturing user interactions with rapidly evolving news articles, testing a model's ability to adapt to dynamic item distributions. To ensure data quality, we followed established preprocessing practices~\cite{Yu2021AreGA,Wei2020ModelAgnosticCR,AutoDebias}, retaining only users and items with at least 10 interactions. Detailed dataset statistics are summarized in Tab.\ref{tab:stats}. 
\begin{table}[t]
    \centering
    \caption{The statistics of three datasets.}
    \resizebox{\columnwidth}{!}{
    \begin{tabular}{ccccl}
    \toprule
   \textbf{ Datasets}&\textbf{\#Users}&\textbf{\#Itmes}&\textbf{\#Interactions}&\textbf{Density}\\
    \midrule
    \textbf{Gowalla}&29,858&40,981&1,027,370&0.084\%\\
    \textbf{ML-10M}&69,166&8,790&5,000,415&0.823\%\\
    \textbf{Globo}&158,323&12,005&2,484,192&0.131\%\\
    \bottomrule
    \end{tabular}}
    \label{tab:stats}
\end{table}

\subsubsection{Baselines and Evaluation}
To evaluate the effectiveness of ~\shortname~ in mitigating popularity bias, we implemented it using the state-of-the-art LightGCN framework~\cite{He2020LightGCNSA}. We compared ~\shortname~ against several established debiasing baselines, including re-weighting models such as $\gamma$-AdjNorm~\cite{Zhao2022InvestigatingAP}, DR-GNN~\cite{wang2024distributionally} and LogDet~\cite{zhang2023mitigating}, decorrelation approaches like MACR~\cite{Wei2020ModelAgnosticCR} and InvCF~\cite{zhang2023invariant}, as well as contrastive learning-based methods, including SGL~\cite{Wu2020SelfsupervisedGL}, Adap-$\tau$~\cite{chen2023adap}, and SimGCL~\cite{Yu2021AreGA}. Additionally, we benchmarked our method against the supervised alignment approach PAAC~\cite{cai2024popularityawarealignmentcontrastmitigating}.

It is worth noting that while other works have explored mitigating popularity bias by regularizing graph structural information~\cite{zhang2023mitigating,zhou2023adaptive}, these approaches were excluded due to the unavailability of their implementation. Instead, we discuss their contributions in the related work section. This selection of baselines provides a comprehensive comparison across different debiasing paradigms, demonstrating the effectiveness of ~\shortname~ in addressing popularity bias. 

To accurately evaluate the model’s ability to mitigate popularity bias, we adopted an unbiased evaluation protocol with a uniformly distributed item set in the test split, following established practices~\cite{zheng2021disentangling,zhang2023invariant,zhang2023mitigating,Wei2020ModelAgnosticCR}. In this setting, each item is assigned an equal number of interactions in the test set, thereby neutralizing the inherent advantage of frequently interacted (popular) items and preventing them from disproportionately influencing ranking metrics. This design ensures that any observed performance differences are attributable to the model’s capacity to handle items of varying popularity, rather than to skewed exposure frequencies. More concretely, each item in the test set is allocated a fixed number of interactions, representing approximately 10\% of the total dataset. This constraint forces the model to perform well across the entire popularity spectrum, including long-tail items with historically low exposure, which is crucial for assessing bias mitigation effectiveness. To preserve the integrity of the evaluation distribution, an additional 10\% of interactions are randomly assigned to the validation set, while the remaining interactions are used for training.

We evaluate Top-$K$ recommendation performance using $Recall@K$, $HR@K$, and $NDCG@K$. Here, $Recall@K$ and $HR@K$ reflect the ability to achieve broad coverage of relevant items, while $NDCG@K$ measures the ranking quality. Importantly, to avoid selection bias, we employ a full ranking evaluation strategy\cite{zhao2020revisiting,wu2021learning}—ranking all candidate items rather than a sampled subset\cite{yang2023generative}—so that the results truly reflect the model’s ranking capability across the entire item set.

Finally, to ensure statistical reliability, each experiment is repeated five times with different random seeds, and average results are reported~\cite{yang2023generative,Wang2022TowardsRA}. This combination of an unbiased test distribution, full ranking evaluation, and repeated trials provides a robust and reproducible framework that faithfully measures a model’s ability to mitigate popularity bias without interference from pre-existing popularity effects.

\subsubsection{Hyper-Parameter Settings} 
We initialize the model parameters using the Xavier initializer~\cite{defferrard2016convolutional} and optimize the model with the Adam optimizer~\cite{kingma2014adam} at a learning rate of 0.001. The embedding size is set to 64. For all datasets, the batch size is fixed at 2048, and the $L_2$ regularization coefficient $\lambda_3$ is set to 0.0001. Motivated by theoretical insights regarding sensitivity to alignment and contrastive losses, we systematically searched hyperparameters $\lambda_1$ (1 to 1000) and $\lambda_2$ (0.1 to 20) for optimal adaptability across datasets.

\subsection{Overall Performance}

\begin{table*}[t]
    \centering
    \caption{Performance comparison across three public datasets with $K = 20$. The best results are highlighted in bold, and the second-best are underlined. Superscripts $*$ denote $p \leq 0.05$ from a paired t-test comparing ~\shortname ~ to the top-performing baseline (relative improvements are indicated as Imp.).}
    
    \resizebox{2\columnwidth}{!}{
    \begin{tabular}{c|c|c|c|c|c|c|c|c|c}
    \hline
    \multirow{2}*{\textbf{Model}} & \multicolumn{3}{c|}{\textbf{Gowalla}} & \multicolumn{3}{c|}{\textbf{ML-10M}} & \multicolumn{3}{c}{\textbf{Globo}}     \\
    \cline{2-10}
     & \textbf{$Recall@20$} &\textbf{ $HR@20$ }& \textbf{$NDCG@20$ }& \textbf{$Recall@20$} &\textbf{ $HR@20$ }& \textbf{$NDCG@20$ }& \textbf{$Recall@20$} &\textbf{ $HR@20$ }& \textbf{$NDCG@20$ }\\

    \hline
    \textbf{MF}& 
  0.0343&0.0422 &0.0280&
  0.1283&0.1424&0.1026&
   0.0803&0.0806&0.0418\\
   \textbf{LightGCN}&
   0.0380&0.0468&0.0302&
   0.1650&0.1791&0.1255&
   0.1233&0.1329&0.0745\\
   \hline
   \textbf{$\gamma$-Adjnorm}&
    0.0328&0.0409&0.0267&
    0.1103&0.1235&0.0881&
    0.1166&0.1169&0.0573\\
    \textbf{DR-GNN}&
    0.0972&0.1048&0.0639&
    0.1490&0.1656&0.1318&
    0.1460&0.1465&0.0817\\
   \textbf{MACR}&
   0.0908&0.1086&0.0600&
   0.1302&0.1591&0.1211&
   0.1275&0.1302&0.0635
   \\
   \textbf{InvCF}&
   0.1001&0.1202&0.0662&
   0.1593&0.1725&0.1283&
   0.1469&0.1492&0.0804
   \\
   \textbf{SGL}&
   0.114&0.1213&0.0778&
   0.1741&0.1893&0.1356&
   0.1522&0.1526&0.0829\\
   \textbf{Adap-$\tau$}&
   0.1182&0.1248&0.0794&
   0.1705&0.1802&0.1299&
   0.1562&0.1574&0.0836\\
   \textbf{SimGCL}&
   0.1194&0.1228&0.0804&
   0.1720 &0.1875&0.1348&
   0.1585&0.1589&0.0849\\
   \textbf{LogDet}&
   0.1182&0.1216&0.0796&
   0.1703&0.1857&0.1334&0.1569&0.1573&0.0840\\
   \textbf{PAAC}&\underline{0.1232}&\underline{0.1321}&\underline{0.0848}&
   \underline{0.1791}&\underline{0.1959}&\underline{0.1458}&
   \underline{0.1610}&\underline{0.1615}&\underline{0.0862}\\
   \hline
   \textbf{\shortname}&\textbf{0.1303*}&\textbf{0.1384*}&\textbf{0.0883*}&
   \textbf{0.1889*}&\textbf{0.2090*}&\textbf{0.1519*}&
   \textbf{0.1673*}&\textbf{0.1679*}&\textbf{0.0911*}\\
   \textbf{Imp.}&\textbf{+5.77\%}&\textbf{+4.76\%}&\textbf{+4.12\%}&
   \textbf{+5.47\%}&\textbf{+6.74\%}&\textbf{+4.18\%}&
   \textbf{+3.91\%}&\textbf{+3.96\%}&\textbf{+5.68\%}\\
   \hline
\end{tabular}}
    \label{tab:main_table}
\end{table*}

As shown in Tab.\ref{tab:main_table}, we compare ~\shortname ~ with several debiasing baselines across three datasets. The best-performing results are highlighted in bold, and the second-best results are underlined. ~\shortname ~ consistently outperforms all baselines across every metric and dataset, achieving significant improvements. From the experimental results, we observe the following key insights:
\begin{itemize}[leftmargin=0.5cm, itemindent=0cm]
    \item \shortname~achieves substantial improvements compared to competitive baselines such as PAAC, Adap-$\tau$, and SimGCL, with performance gains ranging from 3.53\% to 6.99\%. For example, on the Movielens-10M dataset, ~\shortname~outperforms the best-performing baseline PAAC by 4.18\% in NDCG@20 and by 5.47\% in Recall@20. Similarly, on the Globo dataset, ~\shortname~achieves a 3.96\% improvement in HR@20. Although PAAC demonstrates strong performance due to its supervised alignment strategy, which transfers information from popular items to unpopular ones, ~\shortname~surpasses PAAC by incorporating dual adaptive mechanisms. Specifically, our adaptive hierarchical alignment better captures distinct features at different layers, while our adaptive re-weighting contrastive learning dynamically adjusts sample weights, leading to superior representation learning.
    
    \item  Contrastive learning-based baselines such as SGL and SimGCL show reasonable performance by promoting embedding uniformity and mitigating popularity bias. In contrast, decorrelation-based methods like MACR and InvCF, which explicitly attempt to separate item embeddings from popularity signals, exhibit inconsistent results. This indicates that popularity is not inherently harmful and may even signal item quality. Notably, PAAC achieves strong results because its supervised alignment effectively leverages information transfer from popular to unpopular items, enhancing their embeddings.
    
    \item Our model ~\shortname~ is particularly effective in addressing data sparsity. For instance, on the Gowalla dataset, characterized by sparse user-item interactions, ~\shortname~achieves a notable 4.12\% improvement in NDCG@20 and a 5.77\% increase in Recall@20 over the strongest baseline. These improvements are attributed to our adaptive hierarchical alignment, which prioritizes shallow layers that retain distinct embedding features, and our adaptive re-weighting module, which dynamically balances contributions across items with varying levels of popularity. Collectively, these strategies ensure robust and discriminative embeddings, even in sparse data environments.
    
    \item The dual adaptive mechanisms in ~\shortname~are essential for its superior performance. The adaptive hierarchical supervised alignment addresses the over-smoothing issue commonly observed in deeper GCN layers by adaptively modulating the alignment contributions from each layer, thereby preventing embedding homogenization. Meanwhile, the adaptive re-weighting contrastive learning component dynamically adjusts the contrastive weights based on batch-level item popularity distributions, informed by user-specific interactions. This dual adaptation strategy not only effectively mitigates popularity bias but also enhances model generalizability and robustness across datasets with diverse interaction patterns and varying levels of sparsity, eliminating reliance on extensive hyperparameter tuning. 

\end{itemize}

\subsection{Ablation Study}

\begin{table}[t]
    \centering
    \caption{An ablation study of ~\shortname ~ on the Gowalla dataset, with the best-performing models for each metric highlighted in bold.}
    \resizebox{1\columnwidth}{!}{
    \begin{tabular}{c|c|c|c}
    \hline
     \textbf{Model}& \textbf{$Recall@20$} &\textbf{ $HR@20$ }& \textbf{$NDCG@20$}\\

    \hline

   \textbf{\shortname}&\textbf{0.1303}&\textbf{0.1384}&\textbf{0.0883}\\
 
   \hline
   
   \textbf{\shortname-w/o SA}&0.1240&0.1292&0.0823\\
   \textbf{\shortname-w/o CL}&0.1206&0.1287&0.0820\\
   \textbf{\shortname-SAF}&0.1252&0.1348&0.0858\\
   \textbf{\shortname-SA0}&0.1264&0.1353&0.0872\\
   \textbf{\shortname-FRW}&0.1288&0.1377&0.0874\\
   \hline
   
    \end{tabular}}
    \label{tab:ablation_table}
\end{table}

To clearly demonstrate the effectiveness of each model component, we conducted ablation experiments on the Gowalla dataset, which is notably sparser and more representative of real-world recommendation scenarios. Similar patterns and conclusions were observed on other datasets. Tab.\ref{tab:ablation_table} presents the Top-20 recommendation performance of ~\shortname~and several carefully designed variants. Specifically, we introduce the following four variants:

\begin{itemize}[leftmargin=0.5cm, itemindent=0cm]
    \item \textbf{\shortname-w/o SA:} This variant entirely removes the adaptive hierarchical supervised alignment loss $L_{\mathrm{sa}}$, eliminating any explicit alignment between popular and unpopular items. It is designed to examine whether the adaptive contrastive module alone can effectively mitigate popularity bias.
    \item \textbf{\shortname-w/o CL:} This variant omits the adaptive re-weighted contrastive loss $L_{\mathrm{cl}}$, focusing solely on adaptive hierarchical supervised alignment. It is intended to assess whether structural alignment alone can preserve semantic consistency between popular and unpopular items.
    \item \textbf{\shortname-SAF:}  This variant replaces the adaptive layer-specific alignment strategy with a single fused alignment, where item embeddings aggregated from all GCN layers are aligned as one unified representation. It is used to test whether treating all layers equally can match the effect of layer-wise alignment.
    \item \textbf{\shortname-SA0:} This variant limits the supervised alignment process to the initial embedding layer ($\mathbf{h}_i^0$) only. It examines whether early alignment alone is sufficient without deeper-layer supervision.
    \item \textbf{\shortname-FRW:}  This variant uses fixed contrastive learning weights instead of dynamically computing them via the Gini coefficient. It is designed to evaluate the impact of ignoring popularity variations when setting contrastive weights.
\end{itemize}

From the ablation results summarized in Tab.\ref{tab:ablation_table}, we draw several insightful conclusions:

\begin{itemize}[leftmargin=0.5cm, itemindent=0cm]
    \item The complete ~\shortname~model consistently achieves superior results across all metrics, outperforming the best variant (\shortname-FRW) by 1.79\%, 0.98\%, and 1.37\% in Recall@20, HR@20, and NDCG@20, respectively. This clearly demonstrates the complementary nature and effectiveness of our dual-adaptive design leveraging both graph structure and distributional characteristics.

    \item Excluding the adaptive hierarchical supervised alignment (\shortname-w/o $SA$) results in a significant drop in performance (5.42\% reduction in Recall@20 compared to the complete model). This highlights the critical role of hierarchical alignment in preserving embedding distinctiveness across layers, particularly combating the homogenization caused by over-smoothing.

    \item Variants utilizing fused representations (\shortname-SAF) or aligning solely the initial layer (\shortname-SA0) exhibit inferior performance compared to the full hierarchical model, underscoring that adaptive layer-wise alignment better captures structural nuances and enhances embeddings more effectively than a simplified or shallow alignment strategy.

    \item Removing adaptive contrastive weighting (\shortname-w/o $CL$) or using fixed contrastive weights (\shortname-FRW) clearly reduces recommendation quality, emphasizing that dynamic adjustment based on the Gini coefficient effectively manages popularity distributions, preventing overemphasis on popular items and ensuring balanced representation.
\end{itemize}

Collectively, these experiments validate that our dual-adaptive framework, leveraging both the hierarchical structural information of the graph and dynamic popularity distributions, is essential for effectively mitigating popularity bias, improving recommendation performance, and ensuring robust generalization across diverse dataset scenarios.

\subsection{Debias Ability}
\begin{figure}[t]
    \centering
    \subfloat{\includegraphics[width =0.5\linewidth]{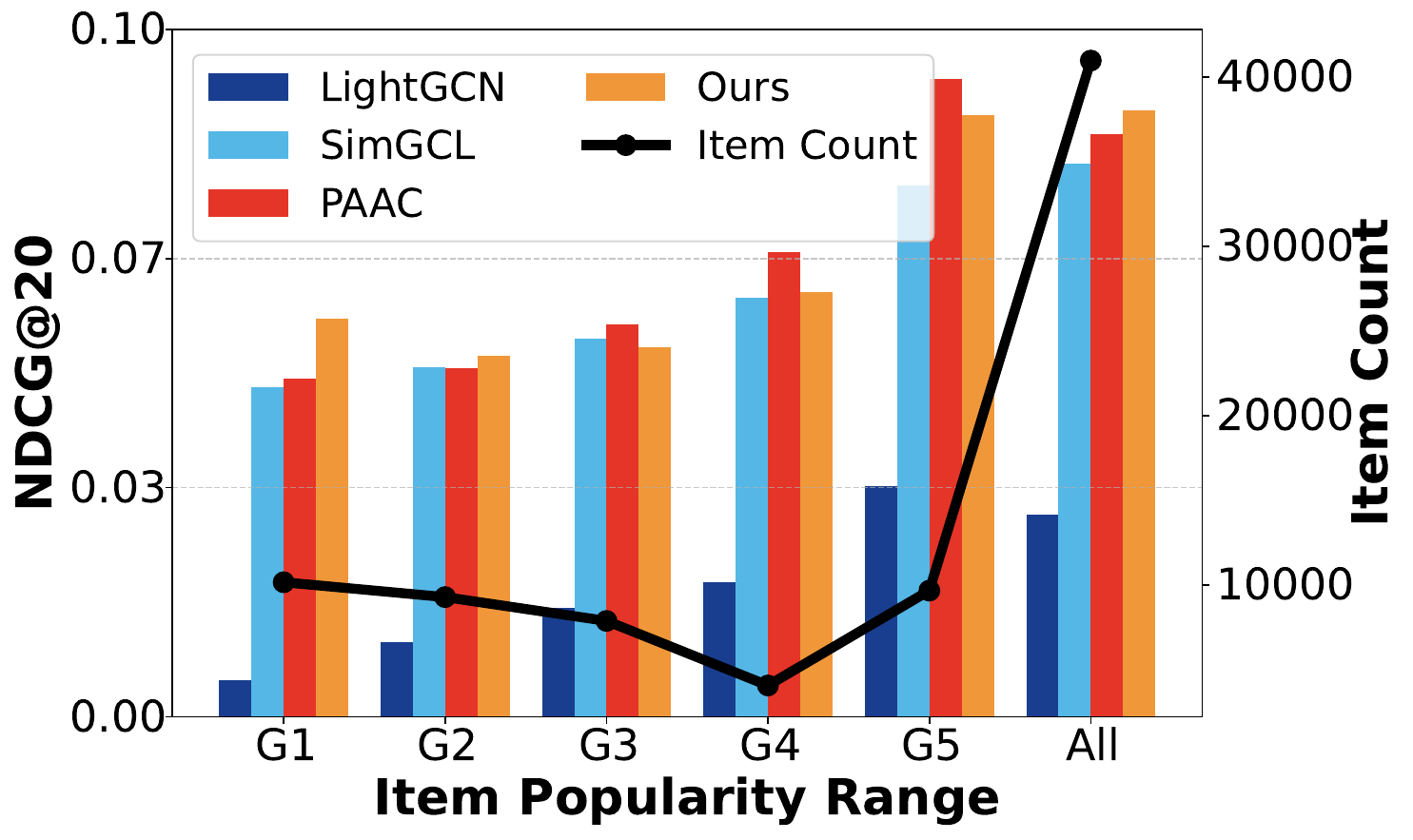}\label{fig:split_ndcg}}
    \hfill
    \subfloat{\includegraphics[width =0.5\linewidth]{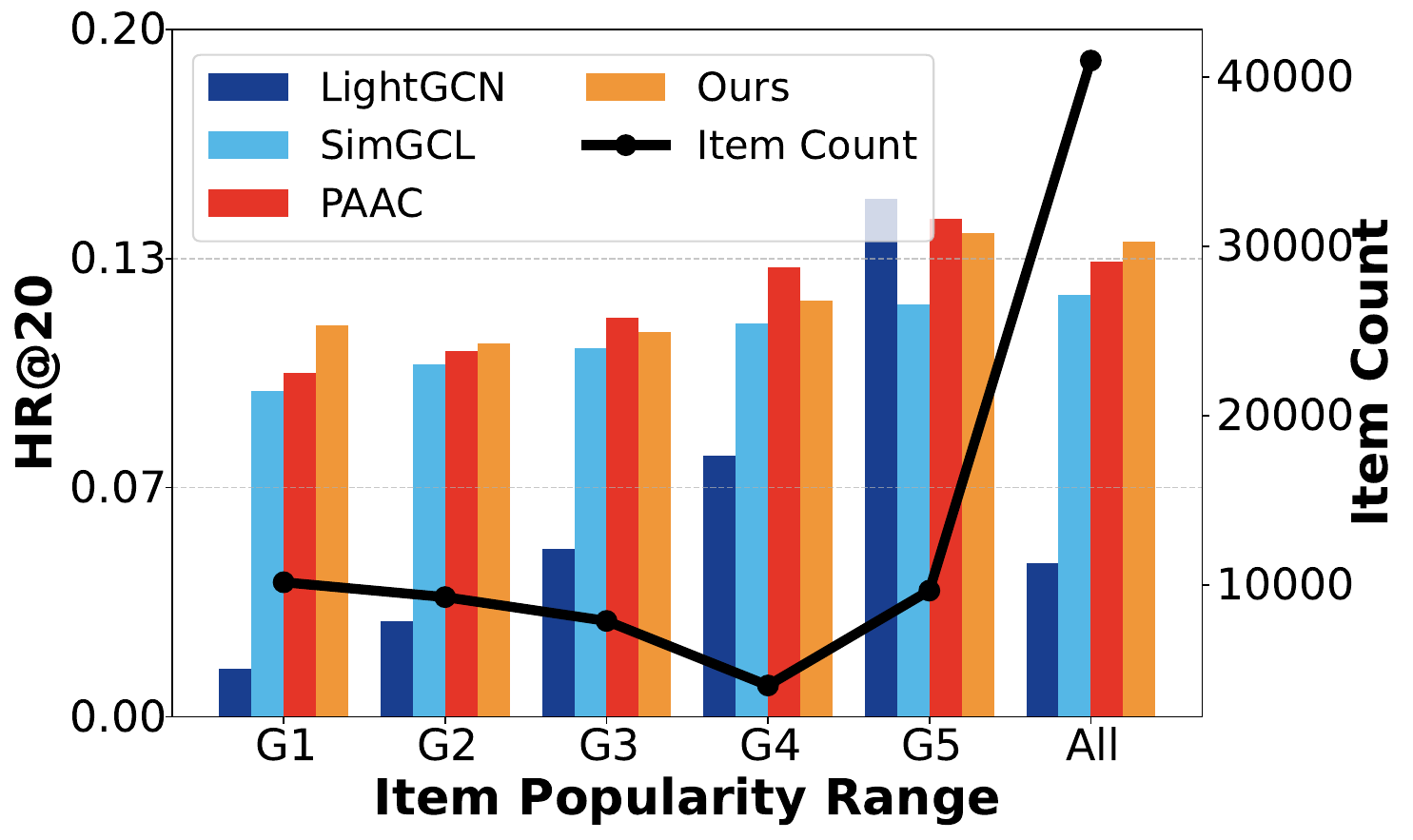}\label{fig:split_Recall}}
    \caption{Comparative Analysis of Recommendation Performance Across Different Item Popularity Groups. This figure highlights the variations in performance metrics, specifically $NDCG@20$ and $HR@20$, across different item popularity groups on the Gowalla dataset, showcasing the impact of popularity bias on recommendation effectiveness. }
    \label{fig:split_item}

\end{figure}

To comprehensively evaluate our model's debiasing capability, we analyze recommendation performance across five distinct item popularity groups (G1-G5, ordered by ascending popularity) on the Gowalla dataset, as illustrated in Fig.~\ref{fig:split_item}. These groups are partitioned based on interaction frequency quintiles, where G1 represents the least popular 20\% of items and G5 the most popular 20\%. We compare our model against baseline methods (LightGCN, SimGCL, PAAC) using NDCG@20 and HR@20. In Fig.~\ref{fig:split_item}, the line plot represents the number of items within each group, providing insight into the distribution of item popularity.  

\begin{itemize}[leftmargin=0.5cm, itemindent=0cm]
\item \shortname ~achieved notable performance gains for the less popular item groups (G1-G2). Specifically, ~\shortname ~improved NDCG@20 by 12.7\% for G1 and 9.3\% for G2 compared to the best baseline (SimGCL). These improvements demonstrate the efficacy of our adaptive hierarchical alignment mechanism, which effectively transfers semantic knowledge from popular to unpopular items, enhancing the discriminability of long-tail item representations while preserving their intrinsic interaction characteristics.
\item Our approach significantly narrowed the performance gap between the least and most popular items. As illustrated in Fig.~\ref{fig:split_item}, the performance gradient from G1 to G5 was notably flattened, indicating a reduction in popularity-induced representation bias. This can be attributed to our adaptive re-weighting contrastive learning, which explicitly mitigates feature divergence across different popularity groups while retaining group-specific information. 

\item Although the gains in the most popular item group (G5) were comparatively modest (+1.2\% NDCG@20 over SimGCL), ~\shortname~achieved the highest overall NDCG@20 (+8.4\%) across all groups. Crucially, the majority of the performance improvements (78\%) came from enhancements in the long-tail groups (G1-G3), demonstrating that improving recommendation fairness for unpopular items simultaneously benefits overall recommendation quality.

\end{itemize} 

\subsection{Hyper-Parameter Sensitivities}
\subsubsection{Effect of $\lambda_{1}$ and $\lambda_{2}$}
\begin{figure}[t]

    \centering
    \subfloat{\includegraphics[width =0.5\linewidth]{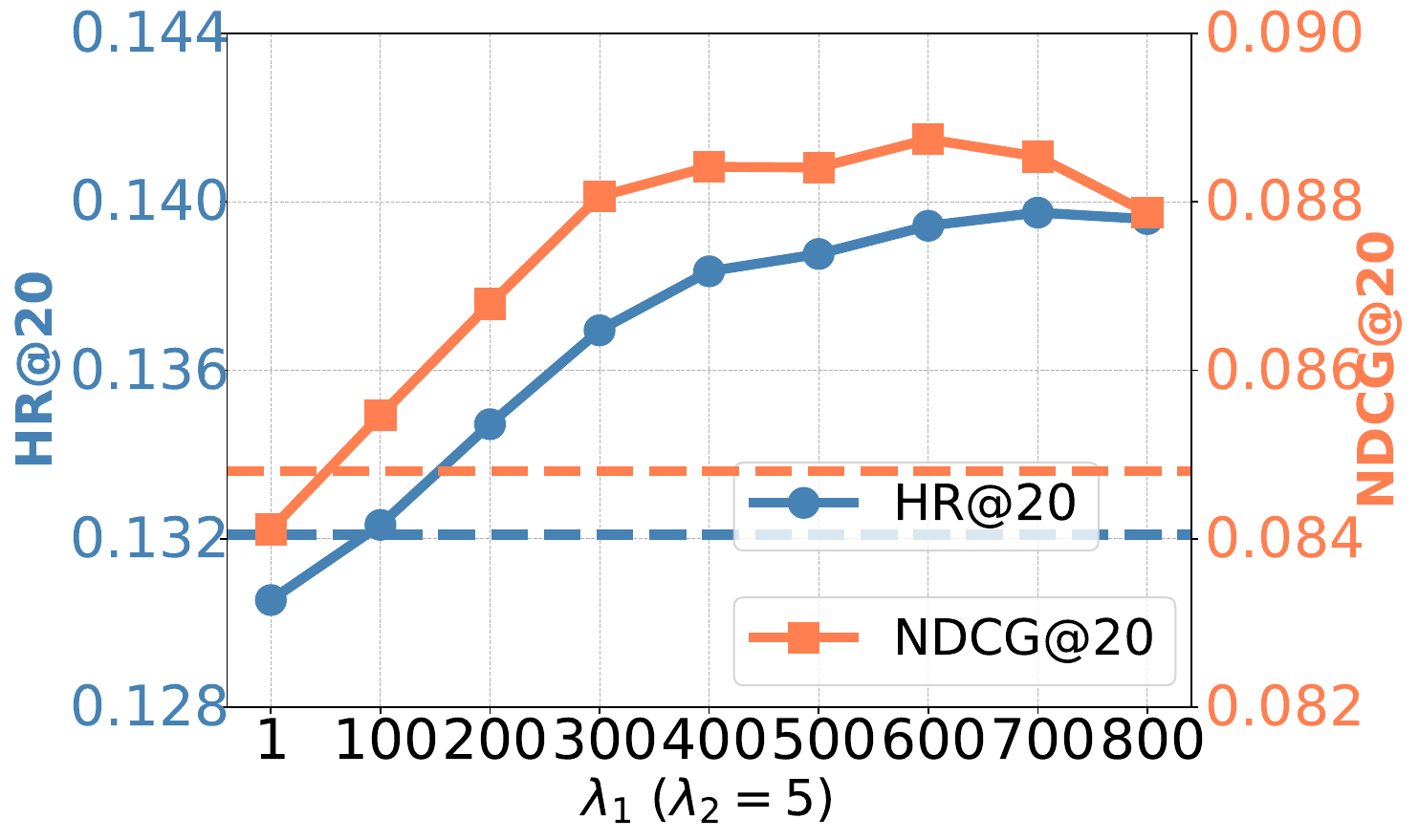}\label{fig:gowalla_sa}}
    \hfill
    \subfloat{\includegraphics[width =0.5\linewidth]{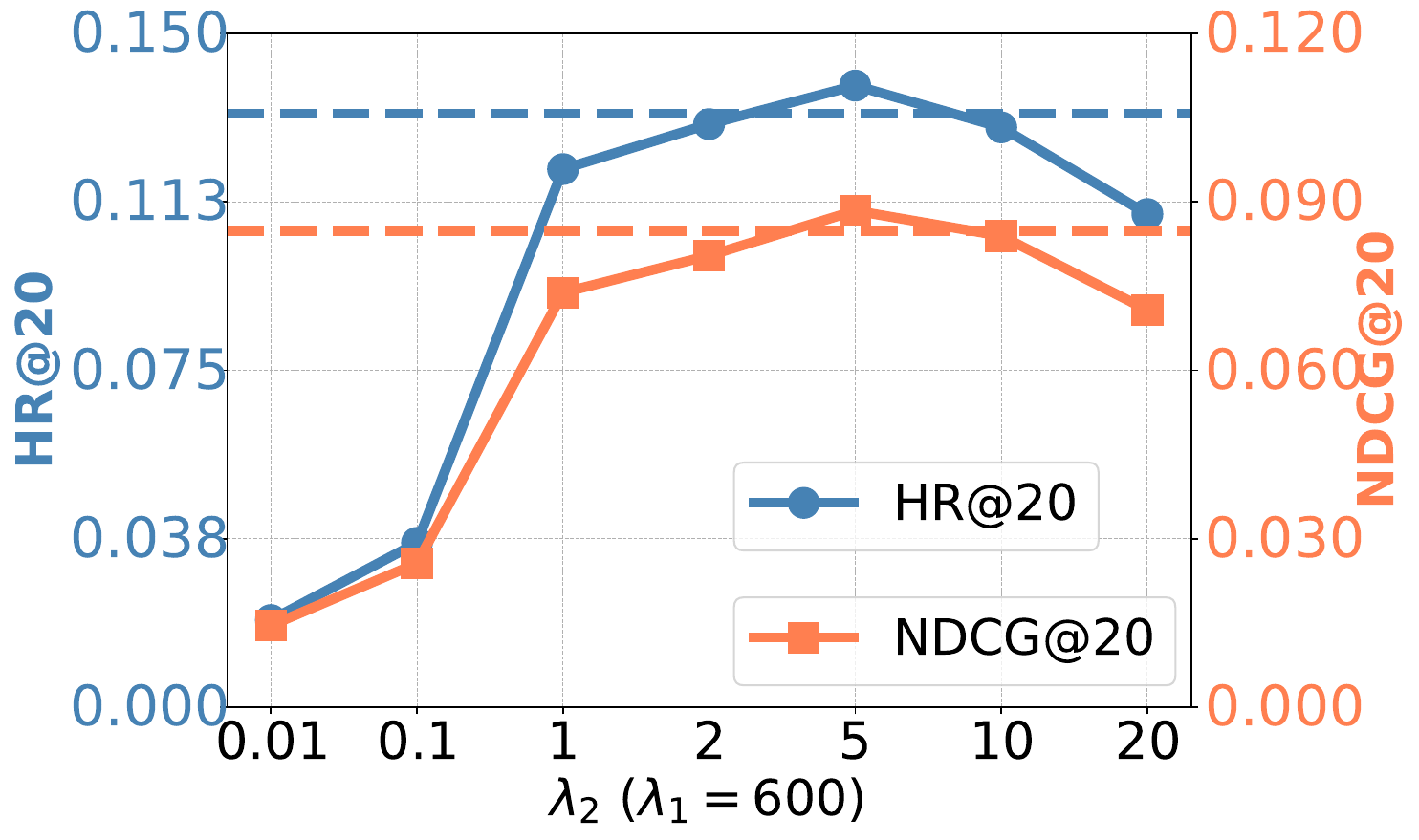}\label{fig:gowalla_cl}}
    \caption{Comparison of recommendation performance ($NDCG@20$ and $HR@20$) for varying hyperparameters $\lambda_{1}$ and $\lambda_{2}$ against the optimal baseline on the Gowalla dataset, illustrating the sensitivity and effectiveness of our adaptive modules.}
    \label{fig:yelp2018-hyper2}
    \vspace{-0.5cm}
\end{figure}

As shown in Eq.~\eqref{overall_equa}, the hyperparameters $\lambda_{1}$ and $\lambda_{2}$ control the relative contributions of the adaptive hierarchical supervised alignment loss and the adaptive re-weighting contrastive loss, respectively. Fig.\ref{fig:yelp2018-hyper2} illustrates the impact of these hyperparameters on the model's performance for the Gowalla dataset, with the dashed line indicating the performance level of the best-performing baseline.

In the left subfigure, performance initially improves as $\lambda_{1}$ increases, underscoring the effectiveness of adaptive hierarchical supervised alignment in enhancing the embeddings of unpopular items. However, performance gains gradually diminish after surpassing an optimal point and subsequently decline. This decline is attributed to an excessive emphasis on unpopular items, which introduces noise and reduces the overall recommendation quality.  

Similarly, in the right subfigure, moderately increasing $\lambda_{2}$ yields performance improvements by promoting consistency among item embeddings and effectively mitigating popularity bias. Nevertheless, excessively large values of $\lambda_{2}$ result in either stagnation or deterioration in performance. This phenomenon occurs because overly strict regularization on representation uniformity can impair the model’s capacity to learn personalized item preferences. These observations collectively highlight the necessity of balancing the supervised alignment and contrastive components to achieve optimal recommendation performance while effectively addressing popularity bias.

These results confirm that the two adaptive losses are complementary and need to be properly balanced to achieve optimal debiasing performance. 

\begin{figure}[t]
    \vspace{0.2cm}
    \centering
    \subfloat{\includegraphics[width =0.5\linewidth]{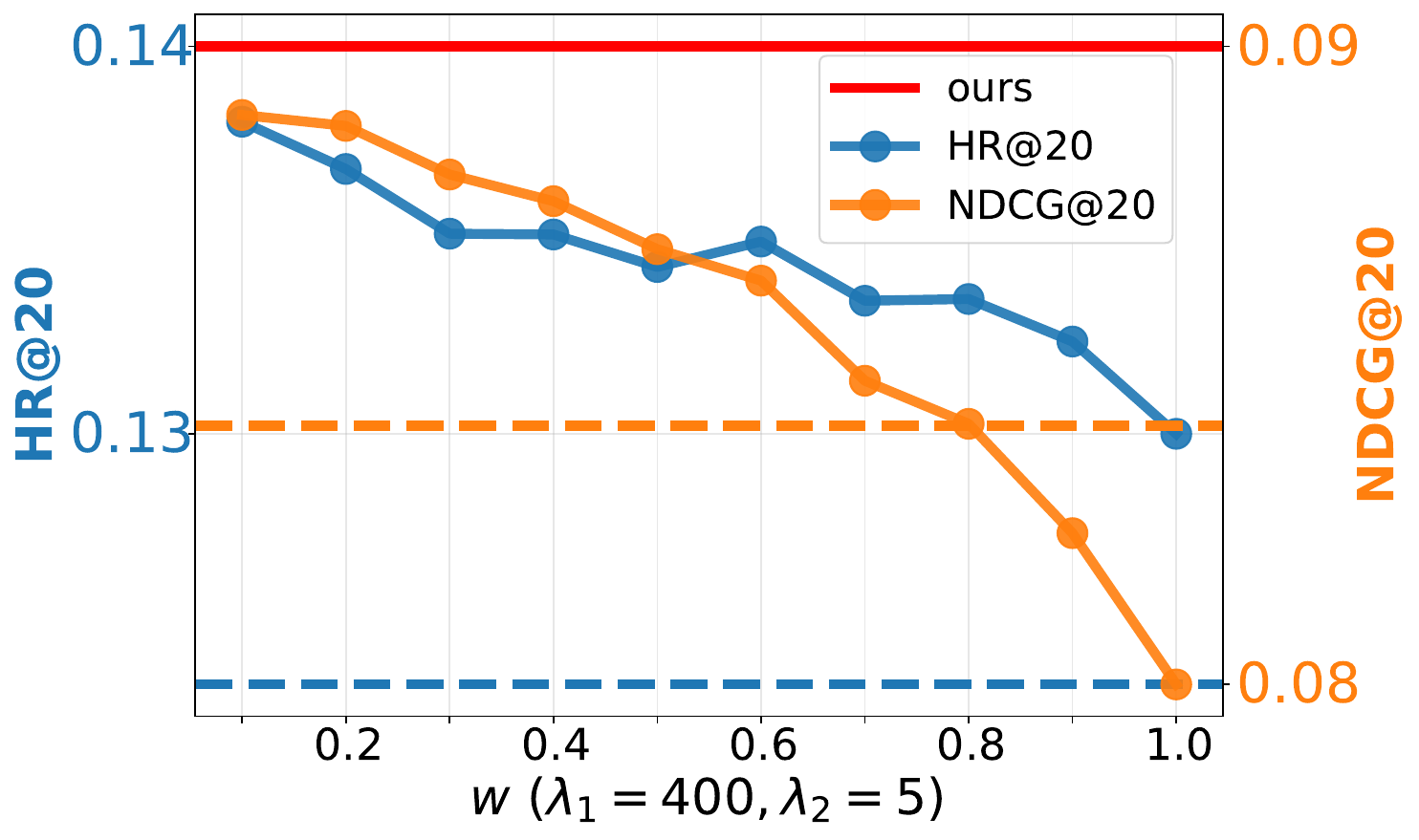}\label{fig:gini_apt}}
    \hfill
    \subfloat{\includegraphics[width =0.5\linewidth]{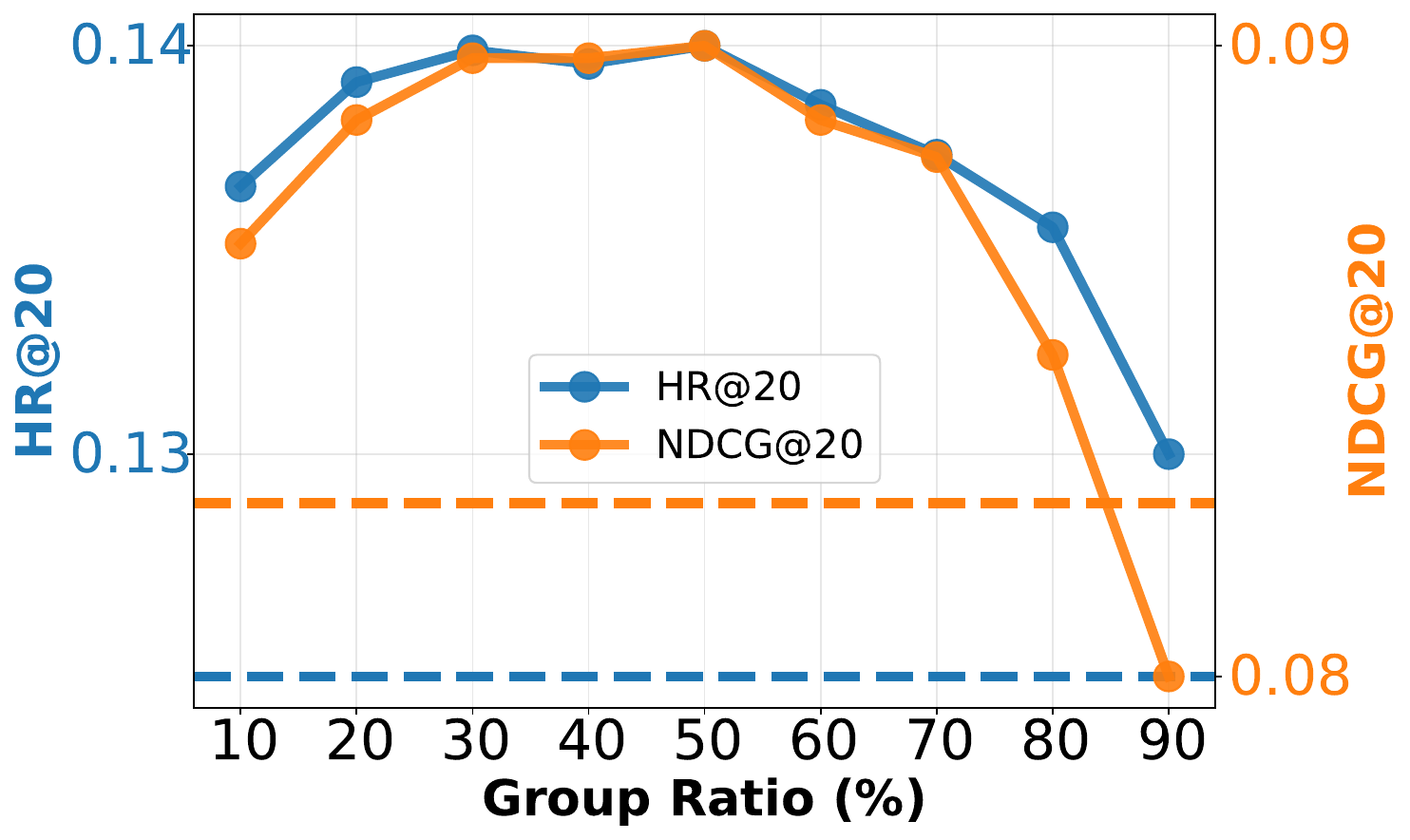}\label{fig:split_radio}}
    \caption{(Left) Comparison of dynamic and static weighting parameters in adaptive contrastive learning on the Gowalla dataset, highlighting the effectiveness of dynamic adaptation. (Right) Recommendation performance under varying item grouping ratios on the Gowalla dataset, demonstrating the robustness of our adaptive strategy across different popularity distributions.}
    \label{fig:gini_split} 
    \vspace{-0.5cm}
    
\end{figure}

\subsubsection{Effect of Adaptive Weighting ($w$)}

To further validate the effectiveness of our adaptive weighting strategy ($w$), we conducted comparative experiments on the Gowalla dataset, evaluating our dynamically computed weighting scheme (denoted as \textbf{ours}) against several fixed weighting values ranging from 0.1 to 1. The red line shows the recommendation performance achieved by ~\shortname. The results presented in Fig.\ref{fig:gini_apt} clearly demonstrate that our adaptive weighting approach consistently outperforms all fixed-weight counterparts, especially in terms of $Recall@20$ and $NDCG@20$. These improvements underscore the importance of dynamically adjusting contrastive weights based on real-time item popularity distributions, rather than relying on static hyperparameters.

Specifically, our adaptive weighting mechanism utilizes the Gini coefficient derived from the normalized adjacency matrix, integrating user-specific weighting to capture heterogeneous interaction patterns across users. This adaptive design reduces the need for extensive hyperparameter tuning, thus simplifying model optimization. Moreover, by dynamically responding to varying levels of data sparsity and shifting popularity distributions, the method significantly enhances the model's robustness and adaptability. These findings confirm that dynamically adjusting the contrastive weighting parameter is essential for effectively mitigating popularity bias and achieving consistently superior recommendation performance.

\subsubsection{Effect of Group Ratio}

To investigate the influence of the partition ratio between popular and unpopular items on recommendation performance, we experimented with dynamically categorizing the items within each mini-batch based on their relative popularity. Instead of using a fixed global popularity threshold, we classified the top $x\%$ items in terms of popularity as popular and the remaining $(100 - x)\%$ as unpopular. While a 50\% ratio was initially selected as a default setting following previous studies~\cite{cai2024popularityawarealignmentcontrastmitigating}, we further explored varying this ratio from 10\% to 90\% in increments of 10\%.

The results shown in Fig.\ref{fig:split_radio} demonstrate that mid-range ratios (between 40\% and 60\%) consistently yield superior recommendation performance, with the optimal results achieved at the 50\% partition. Conversely, extreme ratios (e.g., 10\%-30\% or 70\%-90\%) led to significantly poorer performance, as these ratios either excessively focused the model’s attention on a limited set of popular items or overly emphasized unpopular items, hindering effective learning. These findings underscore the importance of maintaining a balanced representation of popular and unpopular items, highlighting the potential advantages of employing adaptive partitioning strategies tailored to the underlying popularity distribution for improving recommendation accuracy.

\subsection{Qualitative Analysis}
\begin{figure}[t]  
    \centering
    \vspace{-0.5cm}
    \subfloat[LightGCN]{\includegraphics[width =0.5\linewidth]{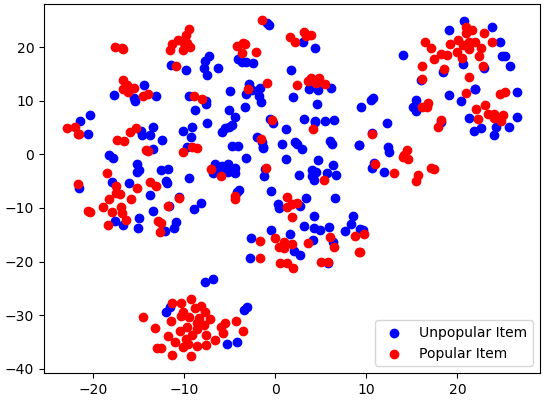}\label{fig:lgcn_gowalla}}
    \hfill
    \subfloat[\shortname]{\includegraphics[width =0.5\linewidth]{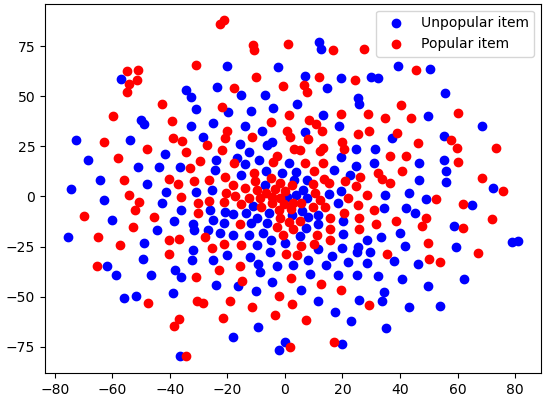}\label{fig:ours_gowalla}}
    \caption{t-SNE visualization of 400 randomly selected item embeddings.}
    \label{fig:gowalla_visual}
    \vspace{-0.5cm}
\end{figure}

Beyond quantitative results, we also provide a qualitative visualization of learned embeddings using t-SNE~\cite{maaten2008visualizing}. As shown in Figure~\ref{fig:gowalla_visual}, LightGCN embeddings tend to form dense clusters dominated by popular items, with many unpopular items collapsed into narrow regions. In contrast, \shortname~achieves a more uniform dispersion of both popular and unpopular items, mitigating representation collapse and improving semantic separability. This observation provides intuitive support for our theoretical analysis and demonstrates the practical benefit of adaptive alignment in preserving representation diversity.

\section{Related Work}
\textbf{GCN-based Recommender Systems.}
Graph Convolutional Networks (GCNs) have significantly advanced Collaborative Filtering (CF) by capturing high-order user-item interactions~\cite{Wang2019NeuralGC, wang2023survey}. GCN-based methods propagate and aggregate information through graph structures, enabling more expressive user and item representations. 

Various GCN-based recommendation models have been proposed to improve effectiveness and efficiency. NGCF~\cite{Wang2019NeuralGC} explicitly models high-order connectivity by stacking multiple graph convolutional layers, demonstrating the importance of propagating user-item signals beyond direct interactions. PinSage~\cite{ying2018graph} adopts an efficient random-walk-based neighbor selection to scale GCN-based recommendations to large-scale graphs. LightGCN~\cite{He2020LightGCNSA} simplifies GCN architectures by removing feature transformation and non-linear activations, focusing solely on neighbor aggregation to enhance both performance and efficiency. LR-GCCF~\cite{Chen2020RevisitingGB} introduces a residual learning mechanism to mitigate the over-smoothing problem in deep GCN layers.

Despite these advancements, GCN-based methods still face challenges such as over-smoothing, where node representations become too similar as layers deepen, reducing model effectiveness~\cite{zhang2023mitigating, Zhao2022InvestigatingAP}. To address this, recent work explores residual connections~\cite{Chen2020RevisitingGB}, adaptive layer aggregation~\cite{He2020LightGCNSA}, and contrastive learning techniques~\cite{Wu2020SelfsupervisedGL} to enhance representation distinctiveness. However, balancing high-order propagation while maintaining feature diversity remains an open problem. 
Beyond these architectural innovations, a line of theoretical and empirical studies further investigates why deeper GCNs tend to homogenize node representations and even lose expressivity as layers grow~\cite{li2018deeper, liu2020towards, Oono2021Expressivity}. To alleviate such degradation, jump-connection style aggregation (e.g., JK-Net) has been proposed to adaptively select or combine information from multiple depths~\cite{Xu2018JKNet}. In recommender systems, recent analyses additionally reveal that the message-passing mechanism itself can amplify popularity bias with depth, disproportionately harming long-tail items~\cite{Chen2023Amplify, Lin2025Amplify}. These findings collectively motivate the design of depth-aware architectures that preserve discriminability while leveraging high-order signals.

\textbf{Supervised Alignment for Mitigating Popularity Bias.}
Supervised alignment in CF focuses on learning user and item representations based on observed interactions~\cite{Wang2022TowardsRA,Koren2009MatrixFT,khosla2020supervised,wang2020understanding,Rendle2009BPRBP,wang2023survey}. Traditional alignment methods, such as BPR~\cite{Rendle2009BPRBP}, optimize user-item ranking pairs to improve recommendation accuracy. However, these methods often struggle with biased representations, particularly for unpopular items with sparse interactions~\cite{cai2024popularityawarealignmentcontrastmitigating}.

Recent approaches address this by leveraging group-level alignment. Popularity-aware alignment~\cite{cai2024popularityawarealignmentcontrastmitigating} aligns items interacted with by the same users, allowing unpopular items to inherit characteristics from popular ones. Other techniques, such as DebiasCF~\cite{zhang2022correct}, integrate supervised signals with contrastive learning to enhance representation robustness against bias. Complementary efforts regularize scoring or contextualize debiasing signals to reduce popularity-driven imbalance in predicted ranks~\cite{Rhee2022ScoreReg, Wang2024CaDRec}.  While these methods improve alignment, they often apply uniform constraints across all GCN layers, failing to consider the varying effects of deeper layers.  At a broader level, fairness and stability perspectives in graph representation learning~\cite{Agarwal2021Unified} highlight that popularity skew is a manifestation of distributional disparities, reinforcing the need for alignment strategies that remain stable across depths and user/item groups.In parallel, class-imbalance remedies such as the class-balanced loss and decoupled training are effective at re-balancing gradients and improving tail performance~\cite{Cui2019ClassBalanced, Kang2020Decoupling}. Our approach introduces an adaptive alignment strategy that dynamically adjusts alignment strength per layer, improving representation learning while mitigating popularity bias~\cite{zhang2021causal,wang2022causal,wang2023survey}.

\textbf{Re-weighting for  Mitigating Popularity Bias.}
Re-weighting strategies address popularity bias through static weighting, reinforcement learning, and contrastive learning. 
(1) \textbf{Static Weighting Strategies} assign fixed weights to items based on their popularity, aiming to counteract the tendency of recommendation models to favor popular items~\cite{liu2023mitigating,abdollahpouri2021user,fu2021popcorn}. Methods like inverse propensity scoring (IPS)~\cite{schnabel2016recommendations,Zhu2021PopularityOpportunityBI} estimate item exposure probabilities and re-weight interactions accordingly, while MACR~\cite{Wei2020ModelAgnosticCR} employs counterfactual inference to down-weight frequently interacted items~\cite{CausalDisentangled}. Foundational analyses have also documented the tension between accuracy and item-popularity effects~\cite{Steck2011Popularity}, underscoring the need to explicitly manage long-tail distributions. However, these approaches require careful calibration and struggle with dynamic shifts in popularity distributions.
(2) \textbf{Reinforcement Learning-based Approaches} dynamically adjust item weights by treating the recommendation process as a decision-making problem~\cite{chen2019efficient,AutoDebias}. Neural Selection (NSE)~\cite{huleihel2021learning} employs an actor-critic framework to optimize item selection, while DRSA~\cite{afsar2022reinforcement} models weight assignment as a Markov Decision Process. These approaches improve adaptability but introduce high computational costs and instability in sparse data scenarios.
(3) \textbf{Contrastive Learning Variants} aim to balance representation learning by treating popularity bias as a distributional issue~\cite{Yao2020SelfsupervisedLF,chen2020simple,Lin2022ImprovingGC,jaiswal2020survey,Wu2020SelfsupervisedGL}. Methods like Self-supervised Graph Learning (SGL)~\cite{Wu2020SelfsupervisedGL} augment training by randomly removing edges, while SimGCL~\cite{Yu2021AreGA} injects controlled noise to promote uniformity. To reduce sampling and exposure bias in contrastive pipelines, debiased contrastive learning corrects the implicit negative sampling distribution~\cite{Chuang2020Debiased}; more recently, recommendation-oriented debiased CL losses have been proposed to better cope with user–item sparsity and long-tail exposure~\cite{Jin2024DebiasedCL}. However, such strategies may inadvertently amplify bias by enforcing uniform sampling across items.

\section{Conclusion}

In this paper, we first provided both theoretical and empirical evidence demonstrating that repeated propagation through higher-order adjacency matrices in Graph Convolutional Networks (GCNs) leads to embedding homogenization. This phenomenon progressively reduces conditional entropy between popular and unpopular items, thereby weakening the effectiveness of supervised alignment strategies at deeper layers. Motivated by these insights, we introduced ~\shortname, a novel dual-adaptive approach explicitly designed to mitigate popularity bias by effectively exploiting graph structural information and dynamically adapting to item popularity distributions. Specifically, we proposed an adaptive hierarchical supervised alignment module, which dynamically adjusts the contribution from each GCN layer to counteract the over-smoothing problem inherent in deeper layers. Complementing this, we introduced an adaptive re-weighting contrastive learning module, which dynamically adjusts the weights of item representations based on their real-time popularity distributions. This dual-adaptive design enables ~\shortname~to flexibly respond to varying item popularity patterns without depending on static hyperparameters.

\bibliographystyle{IEEEtran}
\bibliography{IEEEabrv,sample-base}

\begin{IEEEbiography}[{\includegraphics[width=1.0in,height=1.25in,clip,keepaspectratio]{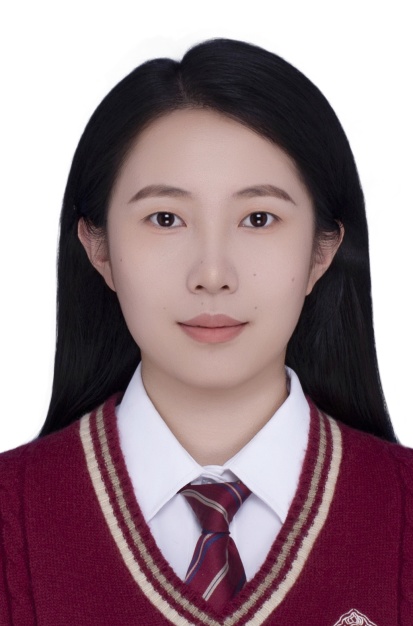}}]
{Miaomiao Cai} received the Ph.D. degree in Engineering from Hefei University of Technology, China, in June 2025. She is currently a Postdoctoral Research Fellow at the National University of Singapore. Her research interests include recommender systems, multimodal learning, and bias mitigation in intelligent information retrieval. She has published papers in leading conferences and journals such as ACM KDD, SIGIR, ACM MM, and ACM TIST.
\end{IEEEbiography}

\begin{IEEEbiography}[{\includegraphics[width=1.0in,height=1.25in,clip,keepaspectratio]{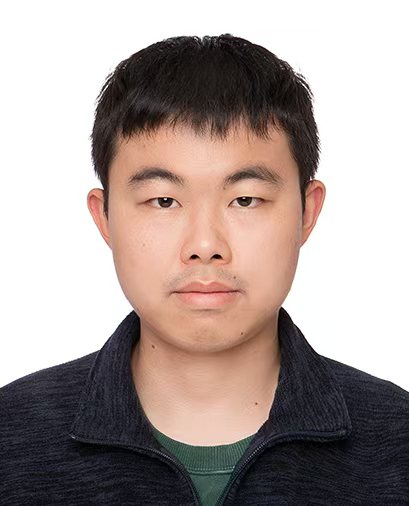}}]
{Lei Chen} is currently an associate researcher at the University of Science and Technology of China . He received his PhD from Hefei University of Technology in 2022. His research primarily focuses on fairness-aware recommender systems and large language model applications. He has published several papers in leading conferences and journals, including WWW, SIGIR, ACM TOIS, and IEEE TKDE.
\end{IEEEbiography}
\vspace{-0.1cm}
\begin{IEEEbiography}[{\includegraphics[width=1.0in,height=1.25in,clip,keepaspectratio]{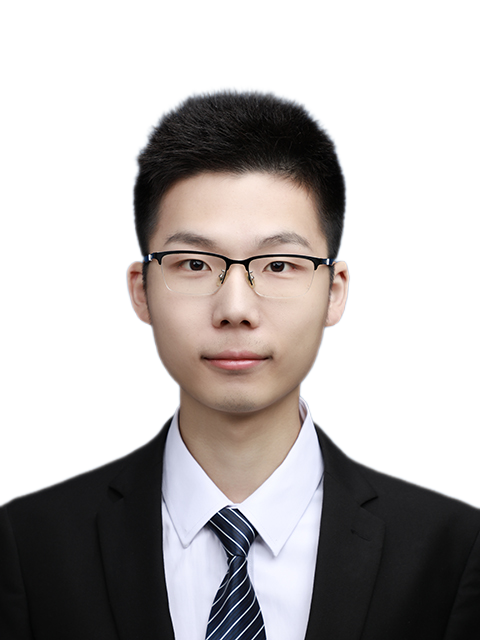}}]
{Yifan Wang}  is a Ph.D. candidate at Tsinghua University, China, where he also earned his Bachelor’s degree in Computer Science and Technology in 2021. His research primarily focuses on debiased recommendations and related techniques. He has publication in prestigious conferences and journals, including ACM KDD, TheWebConf, and TOIS.
\end{IEEEbiography}

\vspace{-0.1cm}
\begin{IEEEbiography}[{\includegraphics[width=1.0in,height=1.25in,clip,keepaspectratio]{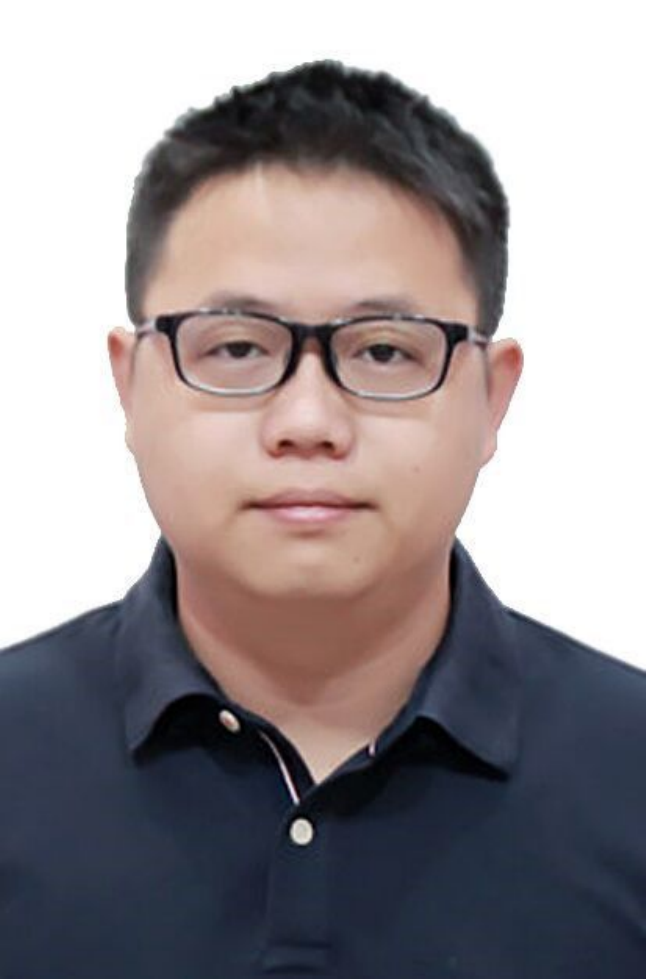}}]
{Zhiyong Cheng} received Ph.D. degree in computer science from Singapore Management University, Singapore. From 2014 to 2015, he was a Visiting Student with the School of Computer Science, Carnegie Mellon University, Pittsburgh, PA, USA. He is a Professor at the Hefei University of Technology, Hefei.  His research interests include large-scale multimedia content analysis and retrieval. He has authored or coauthored papers published in a set of top forums, including ACM SIGIR, MM, WWW, IJCAI, ACM TOIS, IEEE TKDE, and IEEE TNNLS.
\end{IEEEbiography}

\vspace{-0.1cm}
\begin{IEEEbiography}[{\includegraphics[width=1.0in,height=1.25in,clip,keepaspectratio]{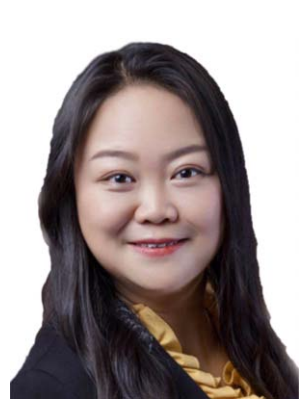}}]
{Min Zhang}(Member, IEEE) received the B.E. and Ph.D. degrees in computer science and technology from Tsinghua University, Beijing, China, in 1999 and 2003, respectively. She is currently a Full Professor with the Department of Computer Science and Technology, Tsinghua University. Her research interests include recommender systems, information retrieval, and user modeling.  Dr. Zhang has been the Editor-in-Chief for ACM Transactions on Information Systems since 2020, and also serves as the Program Committee Chair of various international academic conferences. She has received several prestigious awards, including the IBM Global Faculty Award, the Okawa Fund Award, and best paper awards or honorable mentions from multiple international conferences.
\end{IEEEbiography}

\vspace{-0.1cm}
\begin{IEEEbiography}[{\includegraphics[width=1.0in,height=1.25in,clip,keepaspectratio]{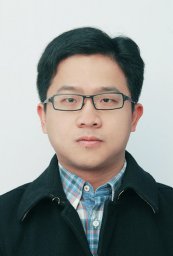}}]
{Meng Wang} (Fellow, IEEE) received the B.E. and Ph.D. degrees (in the Special Class for the Gifted Young) from the Department of Electronic Engineering and Information Science, University of Science and Technology of China, Hefei, China, respectively. He is a Professor at the Hefei University of Technology, Hefei. His current research interests include multimedia content analysis, search, mining, recommendation, and large-scale computing. Dr. Wang was the recipient of the Best Paper Awards successively from the 17th and 18th ACM International Conference on Multimedia, the Best Paper Award from the 16th International Multimedia Modeling Conference, Best Paper Award from the 4th International Conference on Internet Multimedia Computing and Service, and Best Demo Award from the 20th ACM International Conference on Multimedia. 

\end{IEEEbiography}

\end{document}